\documentclass[a4paper]{article}

\usepackage{hyperref}
\usepackage{amsmath,amsfonts,amssymb}
\usepackage{amsthm}
\usepackage{color}
\usepackage{graphicx}
\usepackage{authblk}

\def\ds{\displaystyle}
\def\bea{\begin{array}{c}}
\def\ea{\end{array}}
\def\be{\begin{equation}\bea\ds}
\def\ee{\ea\end{equation}}
\def\bee{\begin{equation}\begin{array}{rcl}\ds}
\def\eee{\end{array}\end{equation}}

\def\Hc{{\mathcal{H}}}

\def\up{|\uparrow\,\rangle}
\def\down{|\downarrow\,\rangle}

\title{Knots and signal transmission in topological quantum devices}

\author{Dmitry Melnikov}

\begin{document}

\maketitle

\vspace{-5cm}
\hfill{ITEP-TH-31/20}
\vspace{5cm}

\vspace{-30pt}
\begin{center}
%\textit{\small $^a$  Department of Physics, Federal University of Paraiba, 58051-900 João Pessoa, PB, Brazil}
%\\ \vspace{6pt}
%\textit{\small $^b$  Department of Theoretical and Experimental Physics, Federal University of Rio Grande do Norte, 59078-970 Natal, RN, Brazil}\\ \vspace{6pt}
\textit{\small International Institute of Physics, Federal University of Rio Grande do Norte, 59078-970 Natal, RN, Brazil}\\ \vspace{6pt}
\textit{\small  Institute for Theoretical and Experimental Physics, B.~Cheremushkinskaya 25, Moscow 117218, Russia}
\\ \vspace{2cm}
\end{center}

\vspace{-2cm}

\begin{abstract}
We discuss the basic problem of signal transmission in quantum mechanics in terms of topological theories. Using the analogy between knot diagrams and quantum amplitudes we calculate the transmission coefficients of the concept topological quantum devices that realize the knot topology. We observe that the problem is in different ways similar to that of transmission on quantum graphs. The desired transmission or filtering properties can be attained by the variation of topology of the device, or an external parameter, which in our model controls the topological phase. One interesting property of the transmission coefficients is the existence of ``self-averaging" phases, in which the value of the transmission coefficient is independent from all the representatives in a chosen family of knots. We briefly discuss physical realizations of the concept devices.
\end{abstract}

%%%%%%%%%%%%%%%%%%%%%%%%%%%%%%%%%%%%%%%%%%%%%%%%%%%%

\section{Introduction}

Topological phases of matter are of great interest for the modern physics research and applications. Topologically protected gapped systems in condensed matter, like those arising in the quantum Hall effect, topological insulators or Majorana fermions, open appealing prospects for design of decoherence resistant quantum information storage. The lack of local dynamics, in the topological phases, implies that the information in such media is stored non-locally, what is a possible quantum way of protecting it against loss.

In classical computation logical operations on information are commonly organized through the control of electric currents flowing through a circuit. The first step from an abstractly defined logical operation to its physical realization can be viewed as solving an appropriate transmission problem. In quantum mechanics this comes down to identifying an appropriate potential and solving the corresponding Sch\"odinger equation. In this paper, we would like to consider such a problem for a class of topological phases.

The peculiarity of the topological theories is in the absence of a local Hamiltonian that would drive the dynamics. Instead of solving the Schr\"odinger problem one should rather focus on a set of global operations external to the model. These are physical operations on the non-dynamical particles coupled to the topological system.  Anyons in the quantum Hall effect (or whatever quasiparticle excitations) can be cited as an example of such particles. They are non-dynamical in the approximation of a large gap between the Landau levels. Physical manipulations (braiding) of anyons are considered as available physical logical operations, at least in principle. These ideas~\cite{Kitaev:1997wr} laid basis for the proposal of a topological quantum computer~\cite{Freedman:2000rc,freedman2003topological}.

In this work we will start with an abstract definition of the non-dynamical particles, as simply Hilbert spaces, where appropriate \emph{topo-logical} operations can act, following a formal approach to the topological quantum computer summarized in~\cite{Melnikov:2017bjb}. Such Hilbert spaces can be provided by representations of the braid group, for example. We will take one known representation and use it to construct quantum circuits, which realize a transmission problem. It is quite well-known that matrix elements of the braid group operators, can be related to the diagrams of three-dimensional knots and links. We will choose a specific connection motivated by a quantum mechanics transmission problem that we will now explain.

In~\cite{drinko2019quantum,drinko2019narrow} a simple model of quantum devices with transmission and reflection properties was considered. In those papers solutions of the free Schr\"odinger equation in one-dimensional spaces of different topologies (graphs) were used to derive non-trivial and sometimes counter-intuitive transmission properties of model quantum devices. One general conclusion that can be drawn from those works, is that even in the system with a free Hamiltonian and most natural boundary conditions, choice of topology allows to design interesting patterns of transmission.

We note that the topological model is in many ways similar to the Schr\"odinger problem on graphs of varying topology. Both graphs and topological theories provide a discretization of space. In the case of graphs it is naturally motivated by convenience in engineering quantum devices. In the case of topological theories it is a direct consequence of non-locality. In both cases the transmission coefficients are oscillating functions of a single parameter: of the particle momentum in the graph model, and of the representation parameter in the topological model. The latter has the interpretation of the coupling constant, when the topological model is described by a Chern-Simons theory. 

The oscillation pattern depends on the choice of the graph topology. For example, combining simple cyclic graphs allows to achieve sequences of peaks of high transmission at certain discrete momenta, separated by gaps of zero transmission~\cite{drinko2019quantum,drinko2019narrow}. In other words, graphs work as filters for all but selected momenta. In the topological model transmission patterns can be simulated if graphs are replaced by knot diagrams, in such a way that the vertices of the graph are replaced by braiding or fusion operations (see figure~\ref{fig:network1} for an illustration). To be precise, the knot diagram is obtained by closing the leads of the graph by an appropriate projection operation. 

Given a knot diagram for a quantum graph the transmission amplitude is expressed in terms of a topological invariant of the knot. Specifically, it is a matrix element of the braiding matrix of the knot diagram, which is expressed in terms of the Jones polynomials of the knot in the given representation. One can further build up the analogy with graphs by noting that cyclic graphs of~\cite{drinko2019quantum,drinko2019narrow} are similar to the $(n,2)$ subfamily of torus knots and links.

Following this approach we study a number of examples of transmission in the topological model and observe that regions of high and low transmission can also be constructed through an appropriate choice of a knot. We focus on the torus and twist knot families and show that the two families have two distinct patterns of transmission. These are characterized by the envelope curves, which show the maximum and the minimum values of transmission for every value of the parameter. 

We find that different families of knots (for example the twist knots, or their generalizations to double braid knots) are characterized by a set of special values of the representation parameter, at which the transmission coefficient is universal for all the family representatives. This property encoded in the properties of the Jones polynomials, which are known to take universal values at some roots of unity (see~\cite{lickorish1986some,murakami1986recursive,przytycki20063} for some known cases). Here we add that distinct families of knots may possess their own universal values of the Jones polynomials. In other words, such values may characterize different knot families.

In this work we do not pursue the exact analogy between the graph problem and the knot realization. The precise patterns of the transmission are different in the two cases. Moreover, there is no unique correspondence between graphs and knot diagrams. However, one aspect of the graph construction make us believe that such a correspondence can be made more precise. Specifically, the set of equations that one uses to solve for the wavefunction amplitudes in the graph model is reminiscent of the famous skein relations of Conway and Alexander that are used to compute topological invariants of knots.

We also discuss possible physical realizations of the proposed circuits and transmission quantum devices based on knots. They can be realized through manipulations of the currents flowing on the edges of the quantum Hall bars. Experimental setup that can control the currents in the necessary way was discussed, for example, in~\cite{barkeshli2014synthetic}. Manipulating edge currents might be conceptually easier than braiding of anyons~\cite{DasSarma:2005zz}, which one commonly imagines when thinking of a topological realization of the quantum computation. Should such a system be constructed in the lab, our knot-based quantum devices would be the basic building block of a topological quantum computer.

The remainder of the paper is organized as follows. In section~\ref{sec:basic} we present the main idea of a quantum transmission problem in terms of knots. In section~\ref{sec:smatrix} we discuss the specific representations of the scattering matrix to be used in the calculation. Section~\ref{sec:graphs} contains a review of the scattering problem on quantum graphs, which served as a motivation for the topological model. We calculate examples of the transmission coefficients for different knots in section~\ref{sec:knots}. Section~\ref{sec:pretzel} considers a generalization of the model to a multichannel scattering. In section~\ref{sec:physical} we discuss possible physical realization of the topological quantum devices. Conclusions are drawn in section~\ref{sec:conclusions}.

%%%%%%%%%%%%%%%%%%%%%%%%%%%%%%%%%%%%%%%%%%%%%%%%%%%

\section{Basic idea of the topological realization}
\label{sec:basic}

Consider a two-level system. Let us denote the state of an electron moving to the right (incident or transmitted electron) as $|\uparrow\,\rangle$ and the state of an electron moving to the left (reflected electron) as $|\downarrow\,\rangle$. A process of electron scattering can be encoded by a unitary matrix $R$, such that
\be
\label{Rmatrix}
R\up \ = \ t \up + r \down\,,
\ee
where $t$ and $r$ are called transmission and reflection coefficients respectively. The matrix $R$ is then
\be
R \ =\ \left(
\begin{array}{cc}
t &  r \\
- \bar r & \bar t
\end{array}
\right)
\ee

Here we will be interested in those $R$ that realize unitary representations of the permutation or braid groups. Braiding is a natural operation on Hilbert spaces of quantum topological theories, for example anyons in quantum Hall effect. In certain representations braid group elements satisfy the following (skein) relations
\begin{eqnarray}
\begin{array}{c}
\includegraphics[width=20pt]{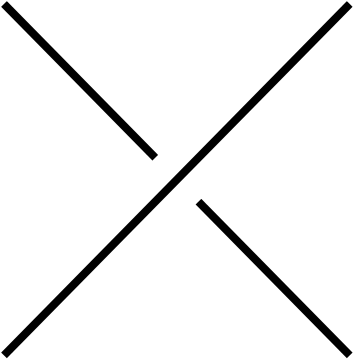}
\end{array}
& = &  A \begin{array}{c}
\includegraphics[width=20pt]{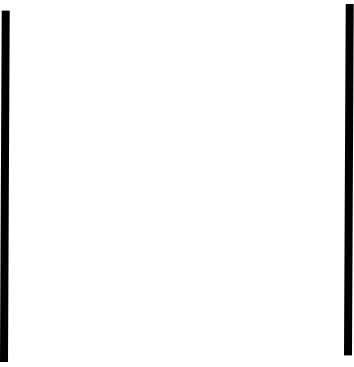}
\end{array} 
+ A^{-1}
 \begin{array}{c}
\includegraphics[width=20pt]{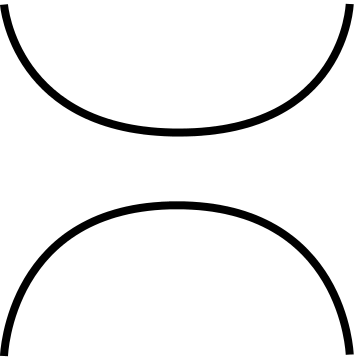}
\end{array}\,, \label{skein1} \\
\begin{array}{c}
\includegraphics[width=20pt]{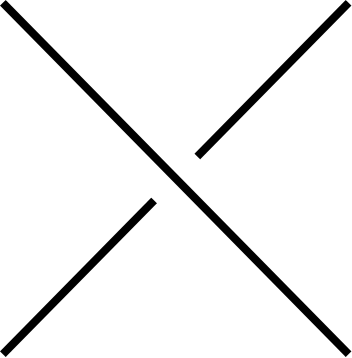}
\end{array}
& = &  A^{-1}
\begin{array}{c}
\includegraphics[width=20pt]{./id.png}
\end{array}
 + A
 \begin{array}{c}
\includegraphics[width=20pt]{./tlieb.png}
\end{array}\,, \label{skein2}
\end{eqnarray}
These relations can be thought as an analog of equation~(\ref{Rmatrix}) for $R$ and $R^{-1}$ respectively. The pair of parallel lines in the right hand side of the equations denote the contribution of transmission to the scattering process, while the cup and cap diagrams correspond to reflection. We also note that relations~(\ref{skein1}) and~(\ref{skein2}) express the connection between the braid group and the Temperley-Lieb algebra.

\begin{figure}[htb]
\begin{minipage}{0.45\linewidth}
 \includegraphics[width=\linewidth]{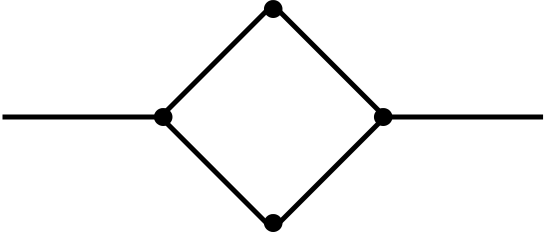}
\end{minipage}
\hfill{
\begin{minipage}{0.45\linewidth}
 \includegraphics[width=\linewidth]{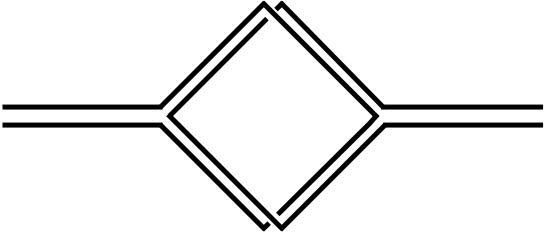}
\end{minipage}
}
\caption{An example of quantum networks considered in~\cite{drinko2019quantum,drinko2019narrow} (left) and a possible topological circuit (right), which corresponds to a specific choice of ``resolutions'' of the vertices. In general, vertices connecting two edges are replaced by braidings and junctions (splitters) by a trivalent vertex. Here the edges will be replaced by double lines.}
\label{fig:network1}
\end{figure}

With this pictorial representation of scattering one can construct networks simulating electron transmitting devices. An example of such a network (graph) is shown on figure~\ref{fig:network1} (left). On the right of the same figure we replace the network by a tangle with a braiding operation inserted at every simple node and a trivalent vertex at every 3-junction. Hence, it is natural to replace the edges of the original diagram by double lines. We will refer to any such choice as a possible ``resolution'' of the network.

\begin{figure}[htb]
\begin{minipage}{0.55\linewidth}
 \includegraphics[width=\linewidth]{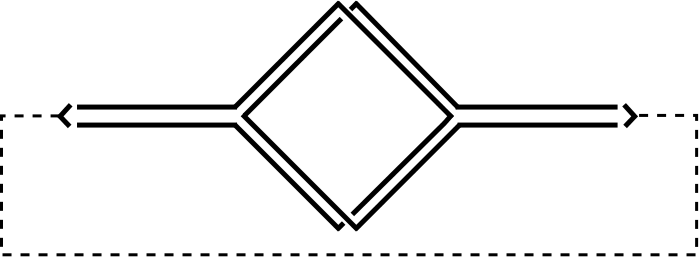}
\end{minipage}
\hfill{
\begin{minipage}{0.35\linewidth}
 \includegraphics[width=\linewidth]{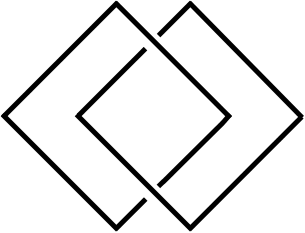}
\end{minipage}
}
\caption{Projecting the result of the scattering on a selected basis state produces a link diagram. Consequently, the scattering amplitude is a topological invariant of the link. In this particular example the link is the Hopf link.}
\label{fig:project}
\end{figure}

As in the case of the scattering problem with stationary Schr\"odinger equation, the graph, or the resolution represents the final state of the scattered electron, which is a linear combination~(\ref{Rmatrix}). The basis of states, corresponding to either $\up$ or $\down$ can be diagramatically presented by
\be
\label{diagbasis}
\up \ = \ \begin{array}{c}
\includegraphics[width=3cm]{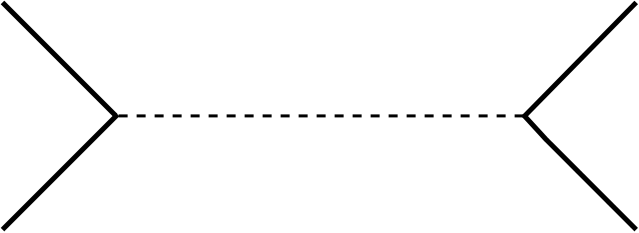}
\end{array},
\qquad\qquad \down \ = \ 
\begin{array}{c}
\includegraphics[width=3cm]{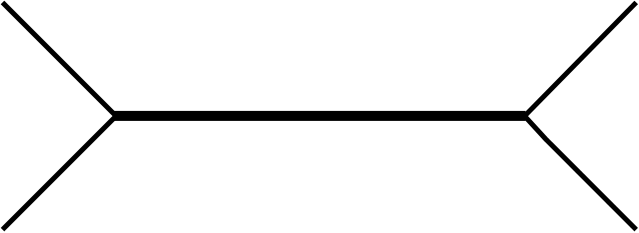}
\end{array}.
\ee
To extract the necessary matrix element the quantum network must be projected on one of these states. Figure~\ref{fig:project} (left) illustrates how the projection, in this model, is done on the $\up$ state, which we choose to be depicted by an intermediate dashed line. 

Connecting this procedure to the technology of calculating topological invariants of knots (see \emph{e.g.}~\cite{Galakhov:2015fna}), we can wipe out the dashed line and replace the amplitude by a diagram of a link, as in figure~\ref{fig:project}.

We note that our discussion is a particular realization of spin networks, originally proposed by Penrose~\cite{penrose1971angular} and developed by other authors in the context of knot theory and quantum information, see e.g.~\cite{Kauffman:2013bh} and references therein.

%%%%%%%%%%%%%%%%%%%%%%%%%%%%%%%%%%%%%%%%%%%%%%%%%%%%%%%

\section{Scattering matrix}
\label{sec:smatrix}

We will consider realizations of the matrix $R$ as an element of a braid group. In this study we will restrict ourselves to the group $B_3$, which is generated by two elements $R_1$ and $R_2$ satisfying
\be
R_1R_2R_1 \ = \ R_2 R_1 R_2\,.
\ee
From the point of view of the graph resolution defined in the previous section, matrices $R_1$ and $R_2$ realize scattering events in different channels of the braid equivalent to the resolution. (We will start considering specific examples in section~\ref{sec:knots} below.) Now we will specify the representation of matrices $R_1$ and $R_2$, that is the space where these matrices act and the explicit form of the matrices.

Define
\begin{eqnarray}
R_1  & = &  A \left(
 \begin{array}{cc}
1 & 0 \\
0 & 1
\end{array}
\right) \ + \ A^{-1}
 \left(
 \begin{array}{cc}
d & 0 \\
0 & 0
\end{array}
\right) , \label{R1}
\\
R_2 & = & A \left(
 \begin{array}{cc}
1 & 0 \\
0 & 1
\end{array}
\right) \ + \ A^{-1}
 \left(
 \begin{array}{cc}
d^{-1} & \sqrt{1-d^{-2}} \\
\sqrt{1-d^{-2}} & d-d^{-1}
\end{array}
\right) \label{R2}
\end{eqnarray}
We note that $R_1$ and $R_2$ are related by a unitary similarity transformation $R_2=FR_1F^{-1}$, where $F$ is a basis changing matrix.
\be
\label{similarity}
F \ = \ \left(
\begin{array}{cc}
 \frac{1}{d} & -\frac{\sqrt{d^2-1}}{d} \\
 \frac{\sqrt{d^2-1}}{d} & \frac{1}{d} \\
\end{array}
\right)
\ee

To complete the definition we specify
\be
\label{d}
A \ = \ {\rm e}^{i\theta}\,, \qquad d \ = \ - A^2 - A^{-2}\ = \ -2\cos2\theta\,.
\ee
Matrices~(\ref{R1}) and~(\ref{R2}) are unitary for the following sets of values of $\theta$~\cite{kauffman2001quantum,Kauffman:2013bh},
\be
\label{eq: domain}
\theta\ \in \ [0,\pi/6]\cup[\pi/3,2\pi/3]\cup[5\pi/6,7\pi/6]\cup[4\pi/3,5\pi/3]\cup[11\pi/6,2\pi]\,.
\ee

It is clear from definitions~(\ref{R1}) and~(\ref{R2}) that the Temperley-Lieb generators are
\be
U_1\left( \begin{array}{c}
\includegraphics[width=20pt]{./tlieb.png}
\end{array}\right) \ = \   \left(
 \begin{array}{cc}
d & 0 \\
0 & 0
\end{array}
\right),
\qquad
U_2\left( \begin{array}{c}
\includegraphics[width=20pt]{./tlieb.png}
\end{array}\right) \ = \   \left(
 \begin{array}{cc}
d^{-1} & \sqrt{1-d^{-2}} \\
\sqrt{1-d^{-2}} & d-d^{-1}
\end{array}
\right).
\ee

The matrices act on a two dimensional space. In fact, this space can be defined by the basis of two diagrams shown in formula~(\ref{diagbasis}). This basis is labeled by the internal line (dashed or thick solid). The scalar product in this space is computed by gluing together the corresponding free legs of the diagrams. Consequently, the above matrices act on the space spanned by $\up$ and $\down$.

Such a space is an example of the space of conformal blocks~\cite{Belavin:1984vu}, labeled by primary operators appearing in the fusion channels of pairs of external primary fields. For example, one might think of diagrams in formula~(\ref{diagbasis}) as two possible ways of fusing spin 1/2 particles, producing either a spin zero particle (dashed line), or a spin one particle (thick solid line). The conformal blocks possess orthogonality properties, so that they indeed can be thought as of a basis. Fixing parameter $\theta$ one fixes the content of the primary fields and possible fusion properties.

Space of conformal blocks and the associated representations of the braid groups are useful in the computation of knot invariants~\cite{Kaul:1991np,RamaDevi:1992np,Ramadevi:1993np}, and this is exactly what we need to compute quantum amplitudes. Before discussing some explicit examples of networks, we will review another quantum network model of signal transmission.

%%%%%%%%%%%%%%%%%%%%%%%%%%%%%%%%%%%%%%%%%%%%%%%%%%%%%%%%

\section{Transmission coefficients in the quantum graph model}
\label{sec:graphs}

In this section we will review the main properties of a different quantum network construction~\cite{drinko2019quantum,drinko2019narrow}. Consider a free Schr\"odinger equation,
\be
H|\Psi\rangle \ = \ E|\Psi\rangle \,, \qquad H \ = \ -\frac{\hbar^2}{2m}\,\frac{d^2}{dx^2}\,,
\ee
on a metric graph~\cite{gnutzmann2006quantum}, like the one shown on figure~\ref{fig:network1} (left). Metric graphs are those with distances associated to edges, with possible exceptions of ``leads'', which are semi-infinite in-coming or out-going edges. Here we will assume all internal edges of the graph to have the same length $\ell$.

At the vertices of the graph one can impose appropriate boundary conditions. Boundary conditions used in~\cite{drinko2019quantum,drinko2019narrow} are the Kirchoff-Neumann ones, implying that the pieces $|\Psi_{ab}\rangle$ of the wave function associated with the edges $E_{ab}$ are glued continuously at the vertices, 
\be
|\Psi_{ab}\rangle \big|_{V_b} \ = \ |\Psi_{bc}\rangle \big|_{V_b}\,, \qquad \forall\, a,b,c\,,
\ee
and that the derivatives satisfy 
\be
\sum\limits_{b} \frac{d}{dx}|\Psi_{ab}\rangle \big|_{V_a} \ = \ 0\,, \qquad \forall\, a\,.
\ee

For a particle with momentum $k$, the transmission amplitude between a lead connected to vertex $V_l$ and a lead at $V_n$ in the graph $A$, can be determined as follows~\cite{andrade2018unitary},
\be
\label{graphtrans}
T_{A;l,n} \ = \ \sum\limits_{a} t_l A_{la}p_{la}^{(n)}\,,
\ee
where $A_{ab}$ is the adjacency matrix of the graph
\be
A_{ab} \ = \ \left\{
\begin{array}{ll}
1\,,     &  \text{$V_a$ and $V_b$ connected}\,, \\
0\,,     & \text{$V_a$ and $V_b$ disconnected, or $a=b$}\,,
\end{array}
\right.
\ee
$p_{la}^{(n)}$ are contributions of the paths from $V_l$ to $V_a$, which can be computed from a set of recursion equations~\cite{andrade2018unitary},
\be\label{paths}
p_{la}^{(n)} \ = \ e^{ik\ell_{la}}\left(p_{al}^{(n)}r_a  + \sum\limits_{b\neq l} t_aA_{ab}p_{ab}^{(n)} + t_n\delta_{an} \right)\,.
\ee
Finally, $\ell_{ab}=\ell$ is the distance between the adjacent vertices and $t_a$ and $r_a$ are the local transmission and reflection amplitudes at every vertex defined by its degree $d_a \ = \ \sum_b A_{ab}$,
\be
\label{localtr}
t_a \ = \ \frac{2}{d_a} \,,\qquad r_a \ = \ \frac{2}{d_a} - 1\,.
\ee

Let us consider a few simplest examples. We will focus on type of graphs considered in~\cite{drinko2019quantum} constructed from degree two and three vertices. The most basic case is a single vertex, $A=0$. In this case, equation~(\ref{paths}) is
\be
p_{11}^{(1)} \ = \ t_1 \ = \ 1 \,,
\ee
that is pure transmission. Note that equation~(\ref{graphtrans}) does not directly apply in this case, because the adjacency matrix is trivial and we used $\ell_{11}=0$.

The next case is two vertices, which has a unique adjacency matrix,
\be
p_{12}^{(2)} \ = \ e^{ik\ell}\left(p_{21}^{(2)} r_2 + t_2 \right) \ = \ e^{ik\ell}\left(e^{ik\ell}\left(p_{12}^{(2)} r_1\right) r_2 + t_2 \right)\,,
\ee
from which one finds
\be
p_{12}^{(2)} \ = \ \frac{t_2e^{ik\ell}}{1-e^{2ik\ell}r_1 r_2}\,, \qquad T_{A;1,2} \ = \ \frac{t_1t_2e^{ik\ell}}{1-e^{2ik\ell}r_1 r_2}\,.
\ee
This is again full transmission, since in the chosen model $r_a=0$ for $d=2$ vertices.

\begin{figure}[htb]
\begin{minipage}{0.45\linewidth}
 \includegraphics[width=\linewidth]{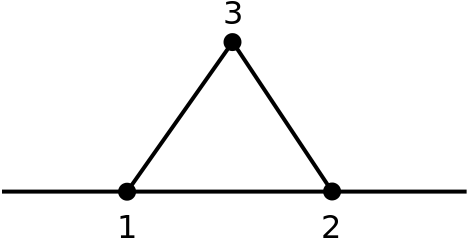}
\end{minipage}
\hfill{
\begin{minipage}{0.45\linewidth}
 \includegraphics[width=\linewidth]{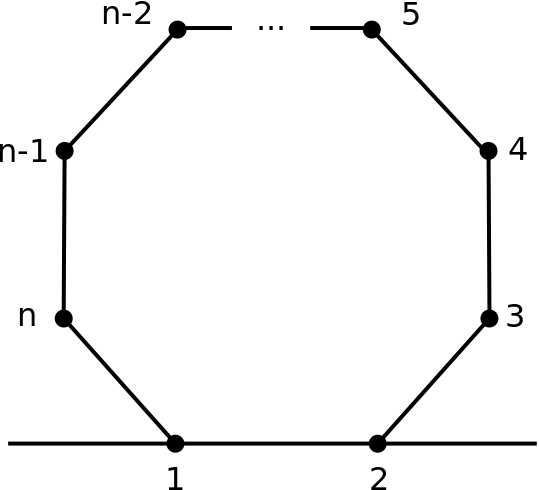}
\end{minipage}
}
\caption{Cyclic quantum graphs were considered in~\cite{drinko2019quantum} as basic building blocks for quantum circuits}
\label{fig:quantgraphs}
\end{figure}

For three vertices there are two choices of a connected graph: all vertices degree two, or two vertices of degree three and one of degree two. One will have full transmission in the first case, while we focus on the second case, illustrated by figure~\ref{fig:quantgraphs} (left).

For this graph we can write the following set of equations~(\ref{paths}):
\begin{eqnarray}
p_{12}^{(2)} \ = \ e^{ik\ell}\left(p_{21}^{(2)}r_2 +  p_{23}^{(2)}t_2 + t_2\right)\,, & \qquad & p_{21}^{(2)} \ = \ e^{ik\ell}\left(p_{12}^{(2)}r_1 + p_{13}^{(2)}t_1\right)\,, \\ 
p_{23}^{(2)} \ = \  e^{ik\ell}\left(p_{32}^{(2)}r_3 +  p_{31}^{(2)}t_3  \right) \,, & \qquad &  p_{32}^{(2)} \ 
= e^{ik\ell}\left(p_{23}^{(2)}r_2 + p_{21}^{(2)}t_2 + t_2\right)\,, \\
p_{13}^{(2)} \ = \ e^{ik\ell}\left(p_{31}^{(2)}r_3 +  p_{32}^{(2)}t_3   \right) \,, & \qquad &  p_{31}^{(2)} \ = \ e^{ik\ell}\left(p_{13}^{(2)}r_1 +  p_{12}^{(2)}t_1 \right)\,.
\end{eqnarray}
Solving for $p_{12}^{(2)}$ and $p_{13}^{(2)}$ one finds the transmission amplitude. The expression is bulky for arbitrary $t_a$ and $r_a$, but for~(\ref{localtr}) it reduces to 
\be
T_{A;1,2} \ = \ \frac{4z(1+2z+2z^2+z^3)}{9+9z+8z^2-z^4-z^5}, \qquad z \ = \ e^{ik\ell}\,.
\ee
In~\cite{drinko2019quantum} a general formula for $n$ cyclically connected vertices was derived, with leads connected to vertices $V_1$ and $V_2$, as in figure~\ref{fig:quantgraphs} (right):
\be
\label{Tcyclic}
T_{A;1,2} \ = \ \frac{4(1-e^{ink\ell})(e^{ik\ell}+e^{i(n-1)k\ell})}{9-e^{2ik\ell}-e^{2i(n-1)k\ell}-8e^{ink\ell}+e^{2ink\ell}}
\ee

For a general cyclic graph, to find the transmission coefficient, one would need to solve a system of $n(n-1)$ equations to determine paths $p_{ab}^{(n)}$. If one ignores the leads, the equations resemble skein relations: they are similar linear relations, which allow to recursively reduce path coefficients to linear functions of themselves, with the coefficients being polynomials of the variable $z=e^{ik\ell}$. The skein relation play the same role in the theory of knot invariants. Consequently we will see similar structure of transmission coefficients in the topological model.

\begin{figure}[htb]
\begin{minipage}{0.45\linewidth}
 \includegraphics[width=\linewidth]{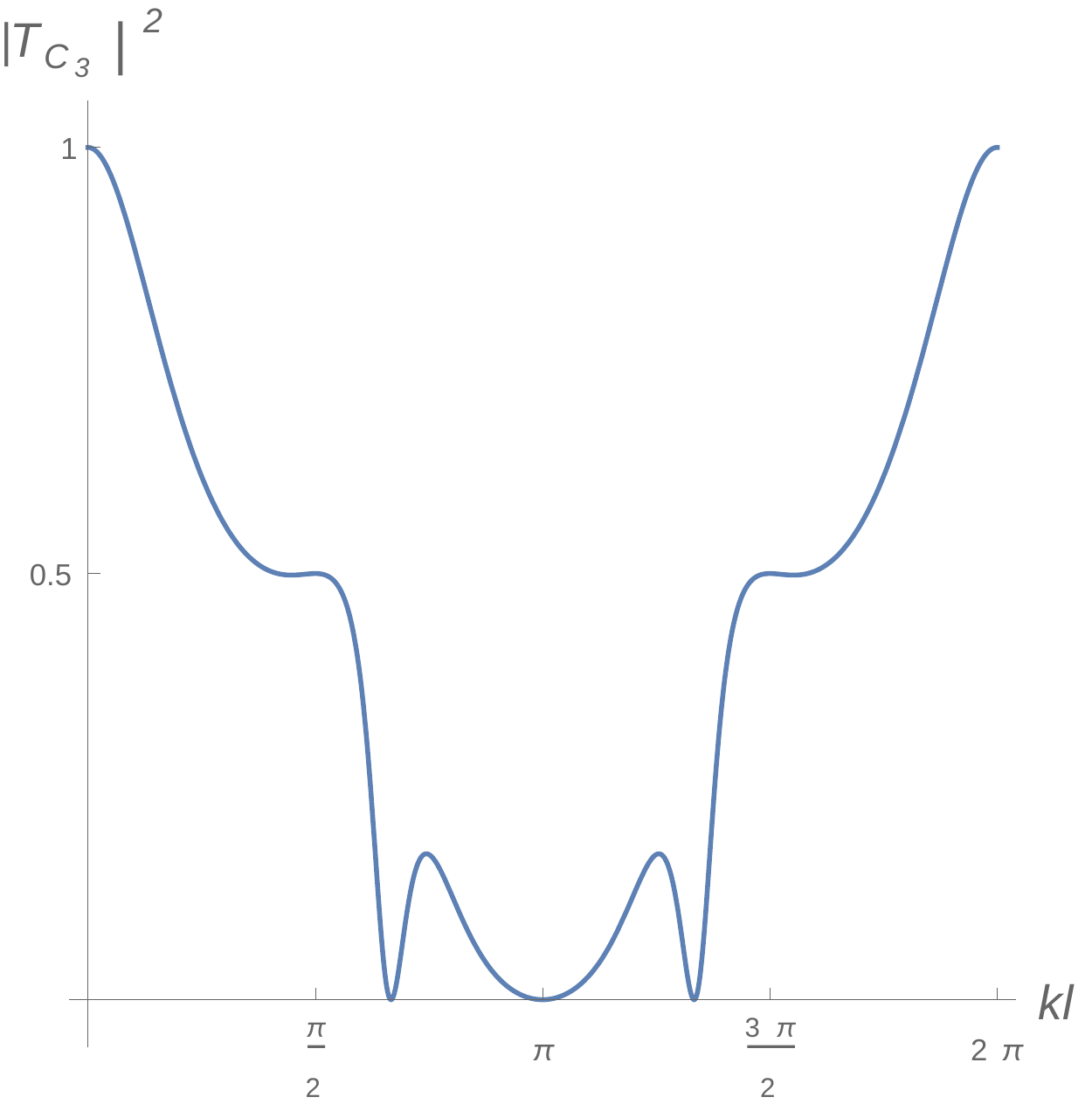}
\end{minipage}
\hfill{
\begin{minipage}{0.45\linewidth}
 \includegraphics[width=\linewidth]{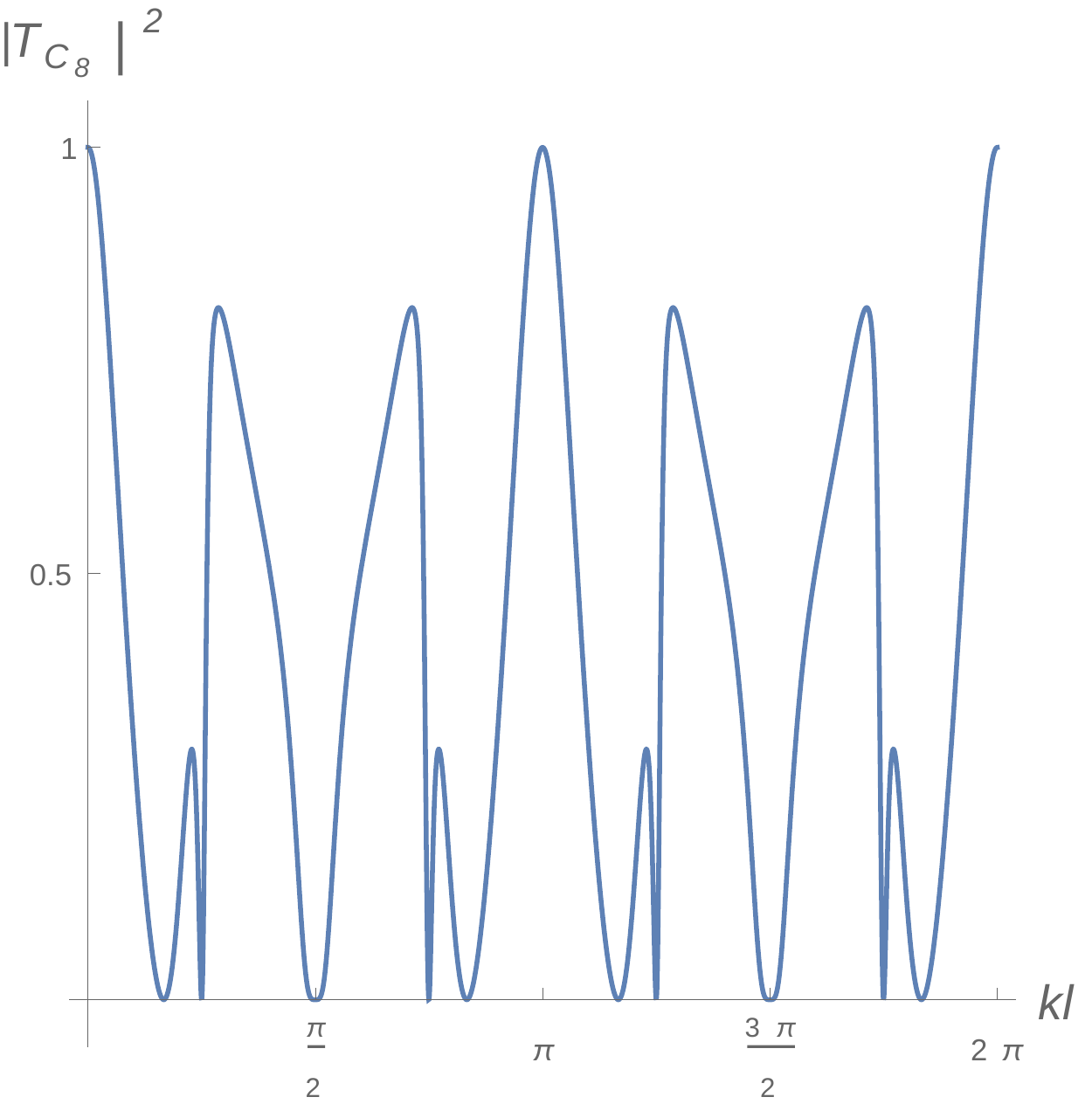}
\end{minipage}
}
\caption{Transmission coefficients of the $C_3$ (left) and $C_8$ (right) quantum graphs.}
\label{fig:Cnplots}
\end{figure}

Let us summarize some of the transmission properties of the quantum graphs. The transmission coefficients of cyclic graphs are in general oscillatory functions of the momentum. The polynomial structure of the numerator defines a set of values of the momentum where the coefficient vanishes. Some of the zeroes of the numerator are not zeroes of the transmission coefficients because of simultaneous zeroes in the denominator. Other possible zeroes of the denominator are irrelevant because they would imply $|T|^2>1$, which is not allowed. In figure~\ref{fig:Cnplots} we show the transmission coefficients of two cyclic graphs, as an example. With growth of $n$ the number of minima and maxima of the transmission coefficients grow.

\begin{figure}[htb]
\begin{minipage}{0.45\linewidth}
 \includegraphics[width=\linewidth]{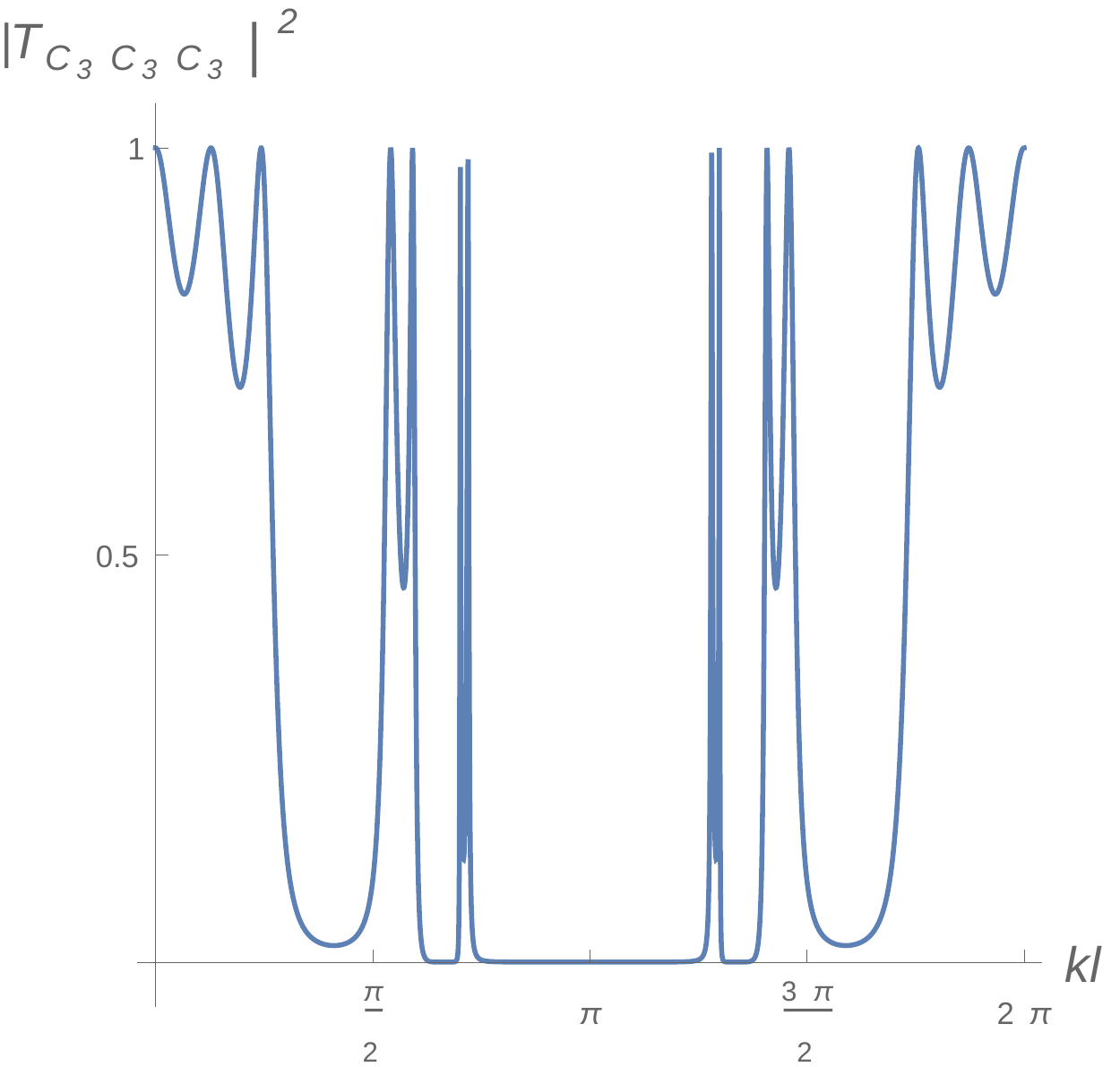}
\end{minipage}
\hfill{
\begin{minipage}{0.45\linewidth}
 \includegraphics[width=\linewidth]{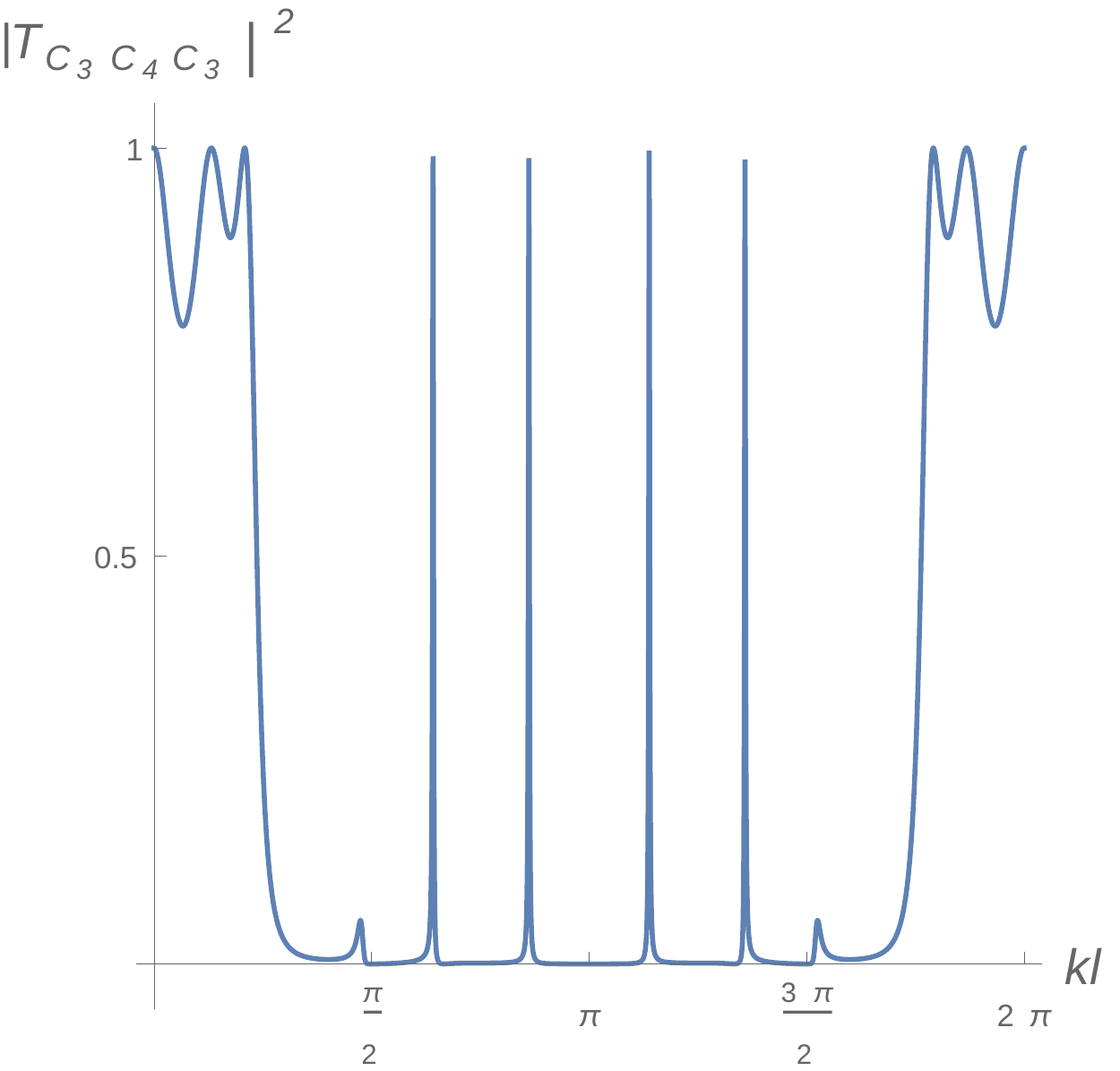}
\end{minipage}
}
\caption{Transmission coefficients for the series connections of the cyclic graphs. Connecting three triangular cycles of figure~\ref{fig:quantgraphs} in series $C_3C_3C_3$ (left) yields regions of zero transmission and peaks of full transmission. If the middle cycle is replaced by a square, $C_3C_4C_3$ (right), one obtains a sequence of sharp separated peaks~\cite{drinko2019quantum}.}
\label{fig:ParCycles}
\end{figure}

Interesting effects happen if one starts combining the cyclic graphs together, using them as basic building blocks. In~\cite{drinko2019quantum} several parallel connections of the cyclic graphs were considered. It was observed that the transmission coefficient in such arrangements exhibits a series of narrow peaks of transmission, which means, such a circuit can serve as a filter allowing only a discrete set of momentum modes. Figure~\ref{fig:ParCycles} shows two examples of the transmission coefficient for cyclic graphs connected in series.

We will now discuss how similar effects of transmission are generated in the topological model considered in this paper.

%%%%%%%%%%%%%%%%%%%%%%%%%%%%%%%%%%%%%%%%%%%%%%%%%%%%%%%%

\section{Transmission coefficients in the topological model}
\label{sec:knots}

In section~\ref{sec:smatrix} we defined the Hilbert state and the explicit representation of the local scattering matrix. Now we will evaluate the transmission amplitudes associated to several families of knots and links, inspired by the quantum graphs discussed in the previous section. First we would like to clarify the connection between the Hilbert space and graph resolutions. In other words, we would like to explain the algorithm of the calculation of the transmission amplitudes for a given resolution.

A knot diagram can be presented as a closure of a braid. For example, the Hopf link (figure~\ref{fig:project}), can be viewed as the following closure of a 4-strand braid:
\be
\label{HopfBraid}
\begin{array}{c}
     \includegraphics[scale=0.2]{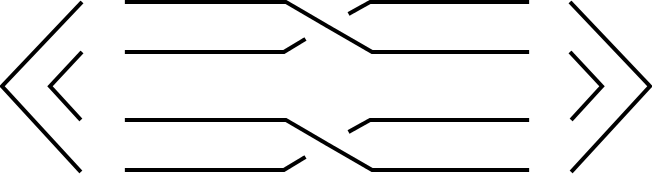}
\end{array} \ = \ \langle \,\uparrow| B \up\,.
\ee
This closure may be understood as an action of a braid group representation, like~(\ref{R1}) and~(\ref{R2}), on the Hilbert space of diagrams~(\ref{diagbasis}). In particular, the bra an ket states used in the closure, are equivalent to the left diagram of equation~(\ref{diagbasis}).

To be more precise, if we would like to understand the above as a scattering process of the initial state $\up$, we should add that the latter is a double line (a composite particle, composed by a fusion of two lines),
\be
\begin{array}{c}
\includegraphics[width=30pt]{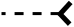} 
\end{array} \ = \ \langle\, 0\,\up\,.
\ee
On the other hand, the braid group is acting on four lines, which actually stands for the double of the Hilbert space, $\Hc\otimes\Hc$. The splitter, as on figure~\ref{fig:network1} encodes the state $\up$ in this bigger Hilbert space,
\be
\label{splitter}
\up \ \to \ 
\begin{array}{c}
\includegraphics[width=20pt,angle=180]{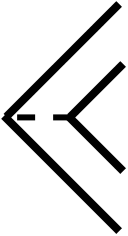}
\end{array}\,.
\ee

Using representation~(\ref{R1}), (\ref{R2}) of braid group $B_3$ (acting on three strands), it is straightforward to construct representations of $B_4$ and higher braid groups. In this section we will restrict ourselves to the case of $B_3$, which is analogous to considering series connections of cyclic graphs. We will use topological equivalence of the closures of the braids (links) to keep all our calculations within the given representation of $B_3$. In particular, the link in equation~(\ref{HopfBraid}) can be replaced by
\be
\label{HopfBraid2}
\begin{array}{c}
     \includegraphics[scale=0.2]{hopfbraid.png}
\end{array} \ = \ \begin{array}{c}
     \includegraphics[scale=0.2]{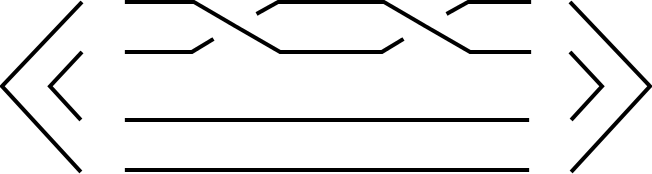}
\end{array}\,.
\ee
In other words, we will use the embedding of $B_3$ into $B_4$, in which the last strand is merely a spectator. The subspace of $\Hc\times\Hc$, where $B_3$ will act corresponds to the first three strands.
 
To match the basis of the representation $(\ref{R1})$ and $(\ref{R2})$ with the diagram basis, we note that from the point of view of the splitter~(\ref{splitter}), action of the braiding on the second and the third line must be diagonal, so we can identify $R_1$ as the generator $b_2\in B_3$. Consequently, $R_2$ is the generator $b_1$, which braids the first and the second strands.

Now we have all the ingredients for the calculations. The amplitude shown in equation~(\ref{HopfBraid2}) is $\langle\,\uparrow |R_2^2\up$, and the corresponding transmission coefficient
\be
\label{t1}
|t_{2_1^2}|^2 \ = \ |\langle\, \uparrow |R_2R_2\up|^2 \ =\  \frac{\cos^24\theta}{\cos^22\theta}\,,
\ee
where the label $2_1^2$ is the Rolfsen classification of the Hopf link~\cite{katlas}. Note that the matrix element is the upper left element of the matrix $R_2^2$. 

\begin{figure}[htb]
    \begin{minipage}{0.45\linewidth}
     \includegraphics[width=\linewidth]{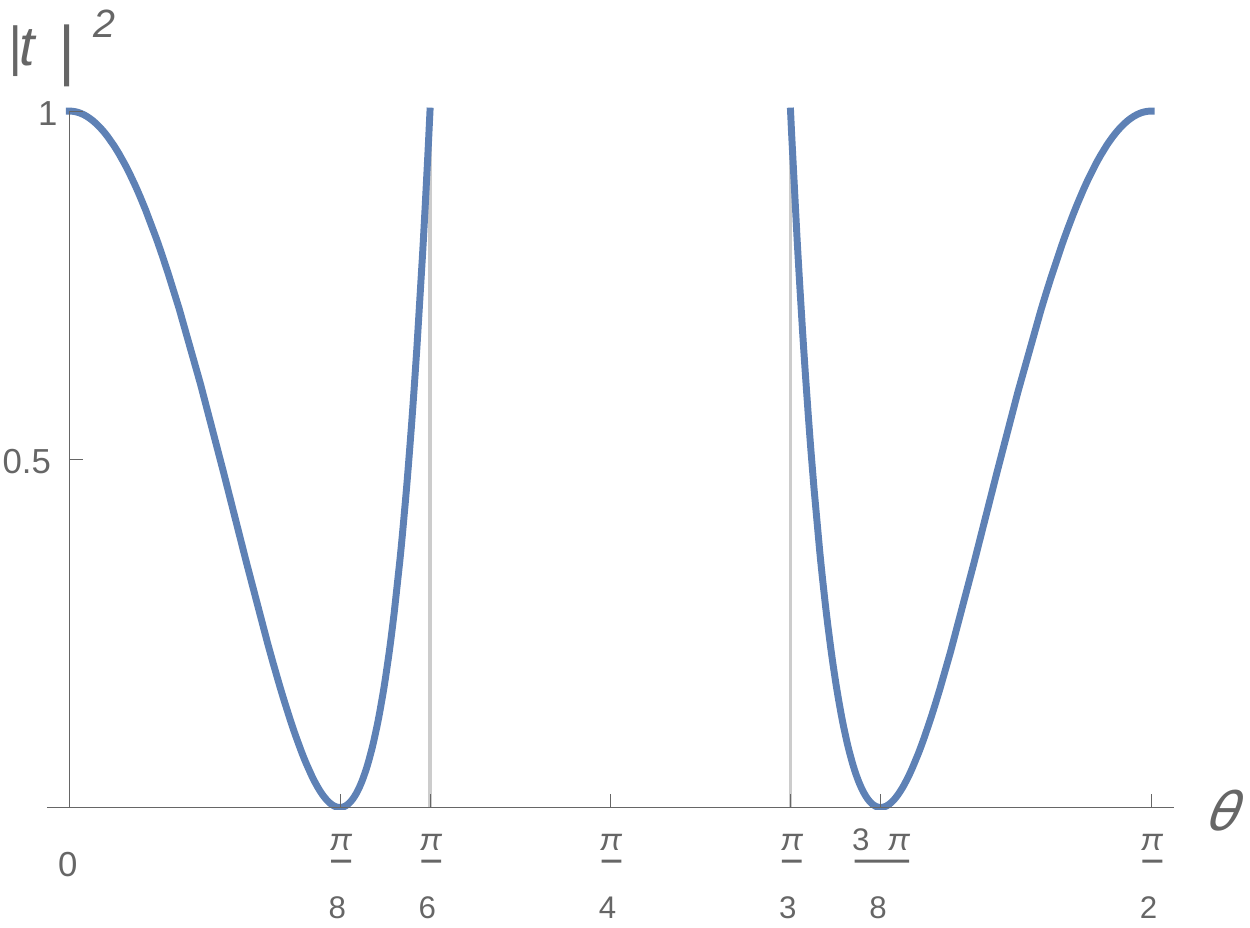}
    \end{minipage}
    \hfill{
    \begin{minipage}{0.45\linewidth}
     \includegraphics[width=\linewidth]{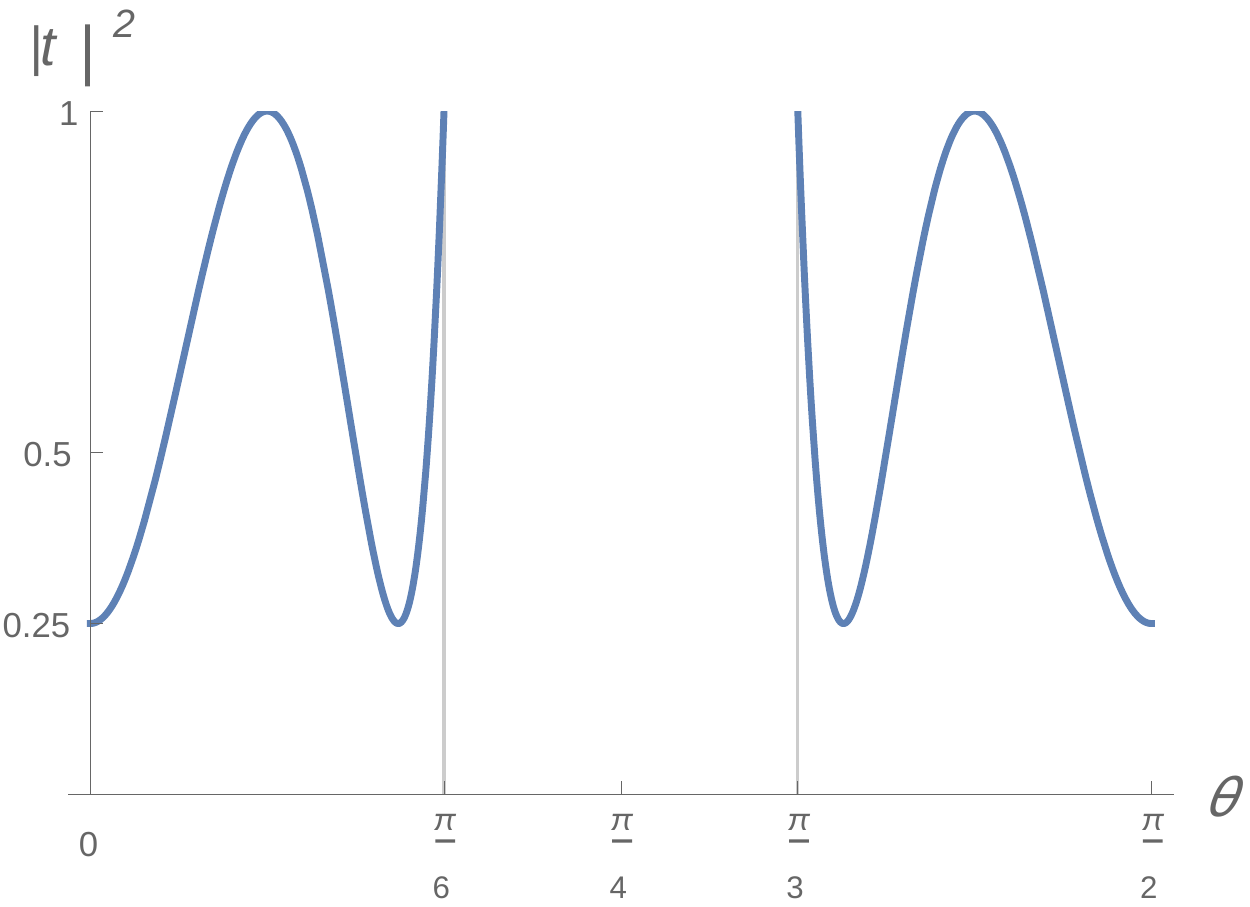}
    \end{minipage}
    }
    \caption{Plots of the transmission coefficients~(\ref{t1}) and~(\ref{eqtrefoil}) associated with the Hopf link (left) and the trefoil knot (right).}
    \label{fig:t1}
\end{figure}

The behavior of this transmission, as a function of $\theta$ is shown in figure~\ref{fig:t1}. It vanishes when $\theta=\pi/4\pm \pi/8\mod \pi/2$. It is maximal $\theta=\pi/2\mod \pi/2$ and $\theta=\pi/4\pm\pi/12\mod \pi/2$. The fact that this coefficient is greater than one for some values of $\theta$ reflects the fact that the representation is not unitary for such values, in agreement with~(\ref{eq: domain}).

{}

We can write the amplitude also as a function of variable $A$,
\be\label{eq:t2}|\langle\,\uparrow|R_2^2\up|=\frac{1+A^8}{A^2+A^6}\ee
and compare it with the Jones polynomial of the Hopf link choosing $A=q^{1/4}$~\cite{katlas}\footnote{In comparison with~\cite{katlas} we will use the polynomials of the mirror images, obtained via $q\to1/q$.},
\be
\label{eq:t3}
Jones(2^2_1)\ =\ - q^{-1/2} - q^{-5/2} \ =  \ -\frac{1+A^8}{A^{10}}\,.
\ee
We can see that it only differs by the denominator, so that
\be
\label{eq:tjhopf}
|\langle\,\uparrow|R_2R_2\up| \ =\ \frac{Jones(2^2_1)}{A^{-6}d}\,.
\ee
Recall that $d$ is defined by equation~(\ref{d}). As in the case of quantum graphs, in the previous section, the role of the denominator is mostly normalization of the result, although in some cases it also removes certain zeroes of the numerator.

In the $A$ variable the plot has periodicity $\pi/2$, so in fact, $q=A^4$ is a more reasonable variable whose degree determines the number of minima of the transmission coefficient, for example.

We will now consider the simplest family of knots and links, which can be seen as an analog of cyclic graphs of the previous section.

\paragraph{Torus links.} The Hopf link is the simples non-trivial example of the family of torus knots and links. In general torus links can be defined as closed curves that can be drawn on the surface of a torus without self-intersections. In the four-strand presentation one can have the $(2,n)$ subfamily characterized by an integer number $n$.\footnote{Torus knots and links are classified by a pair of integer numbers $(m,n)$, which tell how many times the curve winds around two fundamental cycles of the torus. When $m$ and $n$ are coprime, one has a knot. Otherwise it is a link.} For odd $n$ the diagram is a knot. It is a link otherwise. 

The $(2,n)$ subfamily can be constructed by acting $n$ times with the non-diagonal operator $R_2$ on the initial state $\up$. That would generalize the right diagram of equation~(\ref{HopfBraid2}) by considering $n$ crossings instead of two. In the general case the transmission coefficient will be given by
\begin{equation}\label{eq:torustrans}
    |t_n|^2(\theta) \ = \ |\langle\, \uparrow |R_2^n\up|^2 \ .
\end{equation}

To derive $n$ dependence explicitly, we can apply a similarity transformation to diagonalize the matrix $R_2$,
\be
\label{eq:diag} 
R_1 \ = \ F^{-1}R_2F\,,
\ee
where, $F$ is given by~(\ref{similarity}). In other words, we rewrite equation~(\ref{eq:torustrans}) as 
\begin{equation}
    |t_n|^2(\theta) \ = \ |\langle\, \uparrow |(FR_1F^{-1})^n\up|^2 \ =|\langle\, \uparrow |FR_1^nF^{-1}\up|^2 \,,
\end{equation}
which gives a general formula for the transmission coefficient of torus knots and links
\be 
|t_n|^2(\theta) \ = \ |\langle\, \uparrow |FR_1^nF^{-1}\up \ |^2 \ = \  \frac{\sec^4{2\theta}}{16}[1+(1+2\cos{4\theta})^2+(-1)^n(1+2\cos{4\theta})2\cos{4n\theta}]\,.
\ee

The simplest knot representative of the torus family is the trefoil knot, $n=3$. By the above formula the transmission coefficient associated with this knot turns out to be 
\begin{equation}
\label{eqtrefoil}
    |t_3|^2(\theta) \ =\ |\langle\, \uparrow |R_2^3
    \up|^2 \ = \frac{\sec^4{2\theta}}{16}[1+(1+2\cos{4\theta})^2-2(1+2\cos{4\theta})\cos{12\theta}]\,.
\end{equation}

Figure~\ref{fig:t1} (right) illustrates the functional form of the transmission coefficient for the trefoil knot. In comparison with the Hopf link, this coefficient has an additional minimum at $\theta=0\mod \pi/2$. In the meantime there are no zeroes of transmission, $|t_3|\geq 1/4$.

Comparing the amplitude with the Jones polynomial of the trefoil knot,
\be 
Jones(3_1)\ =\ - q^{-4} + q^{-3} + q^{-1} \ = \ \frac{-1+A^4+A^{12}}{A^{16}}\,, 
\ee
we find that
\be
\label{eq:tjtrefoil}
|\langle\, \uparrow |R_2^3 \up| \ = \ \frac{-1+A^4+A^{12}}{A^5+A^9} \  = \ \frac{Jones(3_1)}{(-A^{-3})^3d}\,.
\ee

Jones polynomials of torus knots possess a closed form expression~\cite{Rosso:1993vn,morton1995colored}. 
\be
    Jones(2,n) \ = \ (-1)^{n+1}\frac{q^{-(n+1)/2}}{1+q}\left((-1)^nq^{1-n}+1+q+q^2\right)\,.
\ee
We can therefore express the transmission amplitude for general $n$ in terms of the Jones polynomials
\be
\label{torusamps}
|\langle\, \uparrow |R_2^n \up| \ = \ \frac{(-)^nA^{4}+A^{4n}+A^{4n+4}+A^{4n+8}}{A^{3n}(1+A^4)^2} \ = \ \frac{Jones(2,n)}{(-A)^{3n} d}
\ee
This equation is an analog of formula~(\ref{Tcyclic}) for the cyclic graphs.

\begin{figure}[htb]
    \begin{minipage}{0.45\linewidth}
     \includegraphics[width=\linewidth]{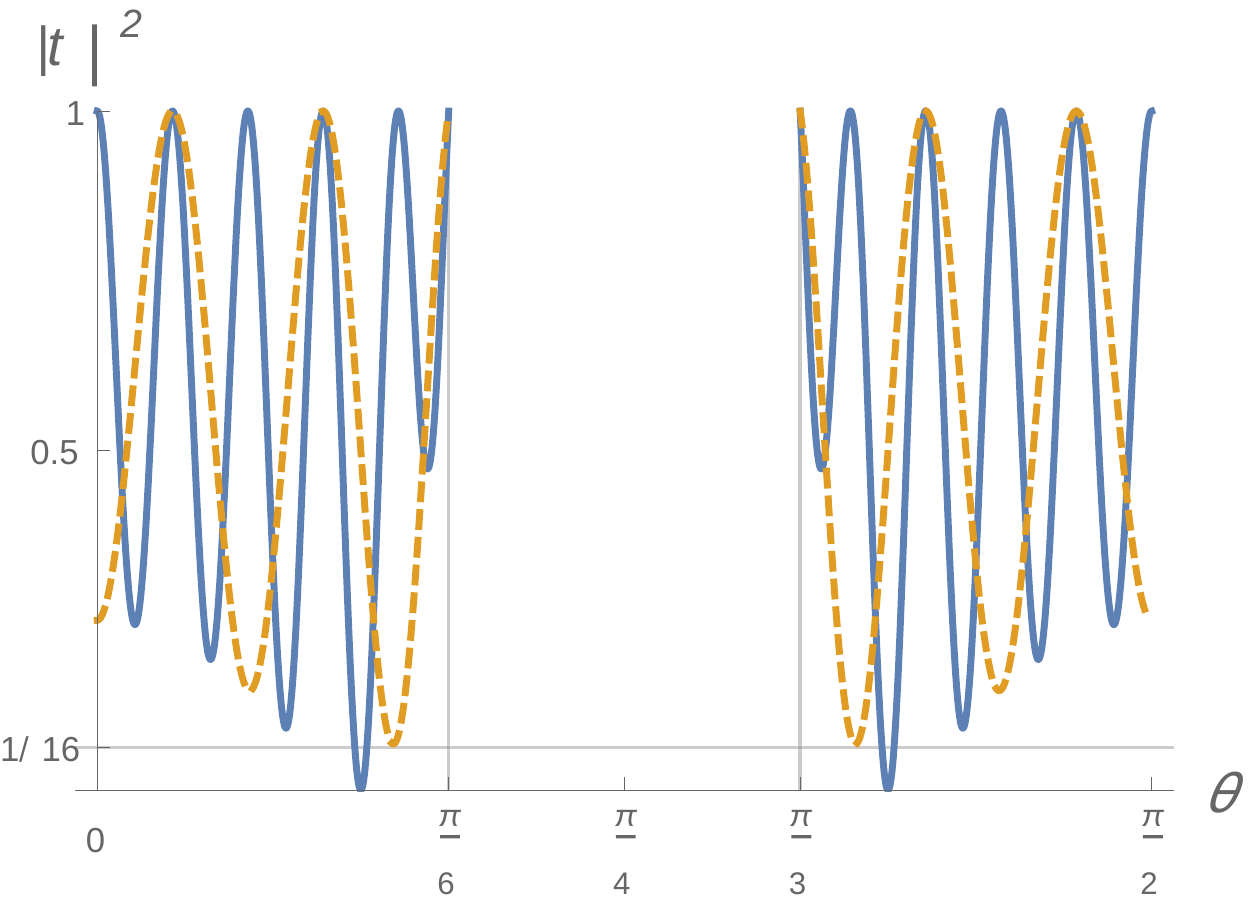}
    \end{minipage}
    \hfill{
    \begin{minipage}{0.45\linewidth}
     \includegraphics[width=\linewidth]{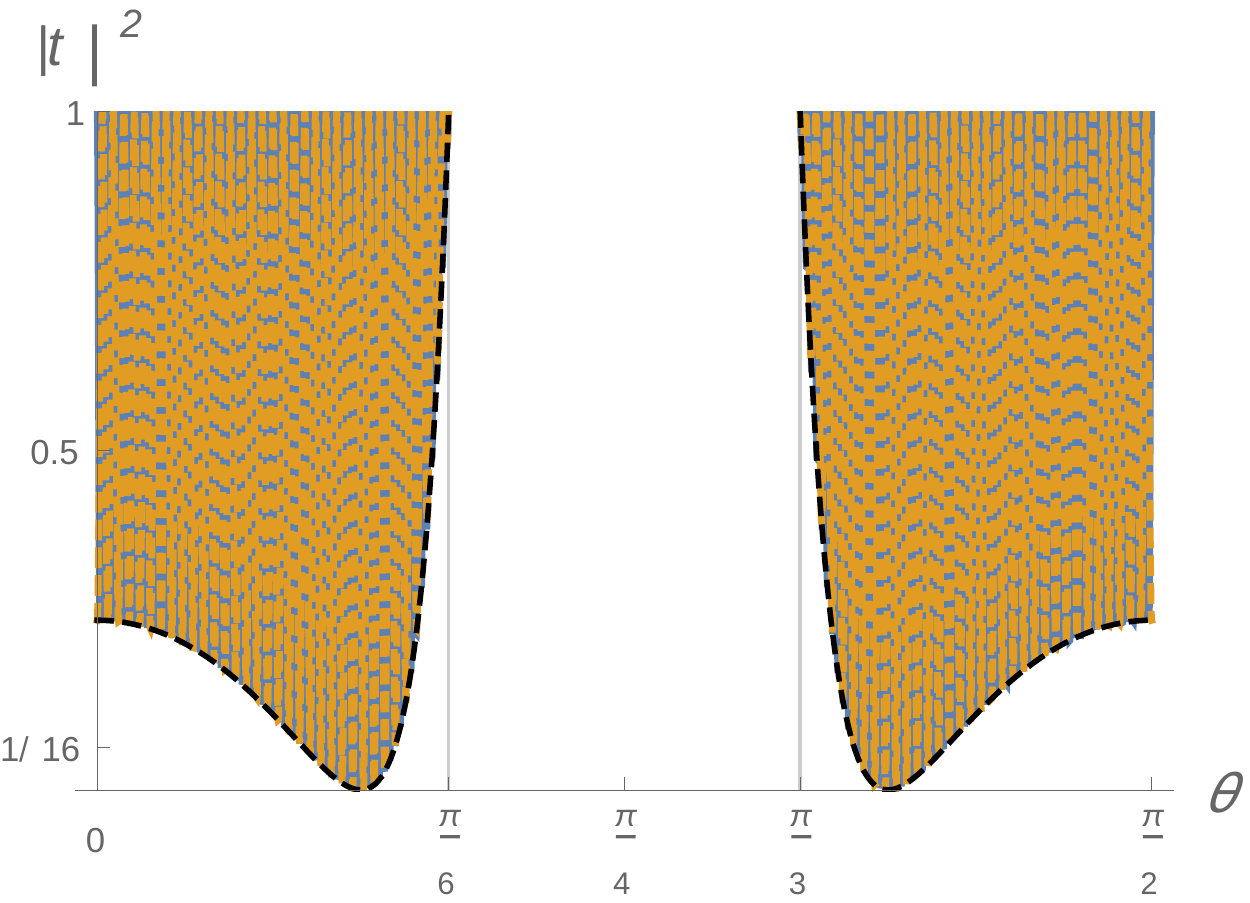}
    \end{minipage}
    }
    \caption{Transmission coefficients of $n=7$ (dashed) and $n=14$ (solid) torus knots (left). Superposition of the coefficients of the $n=100$ and $n=200$ knots (right), which shows envelope curve~(\ref{envelope}).}    \label{fig:tn}
\end{figure}

For general $n$ the transmission amplitudes have zeroes only for $n=2\mod 4$, although for other values the lowest minima asymptote to zero as $n$ becomes large. This is illustrated by figure~\ref{fig:tn}, for example. Indeed, the minima are bound by an envelope curve,
\be
\label{envelope}
\frac{\cos^24\theta}{4\cos^42\theta}\,,
\ee
whose zeroes are at $\theta \ = \ \pi/4\pm\pi/8\mod\pi/2$. These are the zeroes of the Jones polynomials at $n=2\mod 4$. It is not hard to check that these points give maximum transmission for $n=0\mod 4$ and $|t|^2=1/2$ for odd $n$. 

Links (even $n$) have maximal transmission coefficient at $\theta=\pi/2$, while for knots (odd $n$) it is only one quarter. Both knots and link have maximum transmission at the unitarity bound $\theta=\pi/4\pm\pi/12\mod \pi/2$. Hence two envelope curves, (\ref{envelope}) and $|t|^2=1$ give the maximum and the minimum possible values of the transmission coefficients for the torus family $(2,n)$ for each value of $\theta$.

\paragraph{Networks.} After introducing a simple family of knots and links one can move to less regular examples. We would like to think of such examples as combinations of more simple building blocks into arbitrary networks. In the simplest approach adopted in this paper, the connection of basic blocks can be realized through the braiding in the intermediate channel, through application of $R_1$. In the present scenario this allows to generate amplitudes of the form
\be
\begin{array}{c}
\label{twistnetwork}
     \includegraphics[scale=0.25]{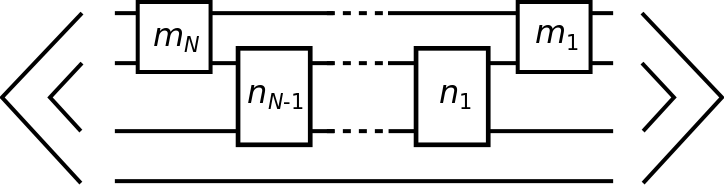} 
\end{array} \ = \ \langle\, \uparrow |R_2^{m_N}R_1^{n_{N-1}}\cdots R_1^{n_2} R_2^{m_2} R_1^{n_1} R_2^{m_1}\up 
\ee
which connects $N$ blocks of $R_2^{m_k}$ intertwined by $N-1$ blocks of $R_1^{n_j}$. Note that numbers $m_k$ and $n_j$ can be both positive and negative. 

One interesting property of such chain networks can be observed for $n_j=1$ and $m_k=m\leq 3$. The corresponding transmission coefficients of such chains are either the same as the ones of the basic blocks $R_2^m$ or unity. For example,
\be
|\langle\, \uparrow |R_2^{3}R_1\cdots R_1 R_2^{3} R_1 R_2^{3}\up|^2 \ = \ \left\{
\begin{array}{ll}
    |\langle\, \uparrow |R_2^{3}\up|^2\, & N=1,2\mod 3\,, \\
    1\,, & N=0 \mod 3\,, 
\end{array}
\right.
\ee
where again, $N$ is the number of $R_2^m$ blocks. This property implies that Jones polynomials of the corresponding links agree up to the framing and mirror reflections, although it does not trivially follow from the braid group itself.

\begin{figure}[htb]
    \begin{minipage}{0.45\linewidth}
    \includegraphics[width=\linewidth]{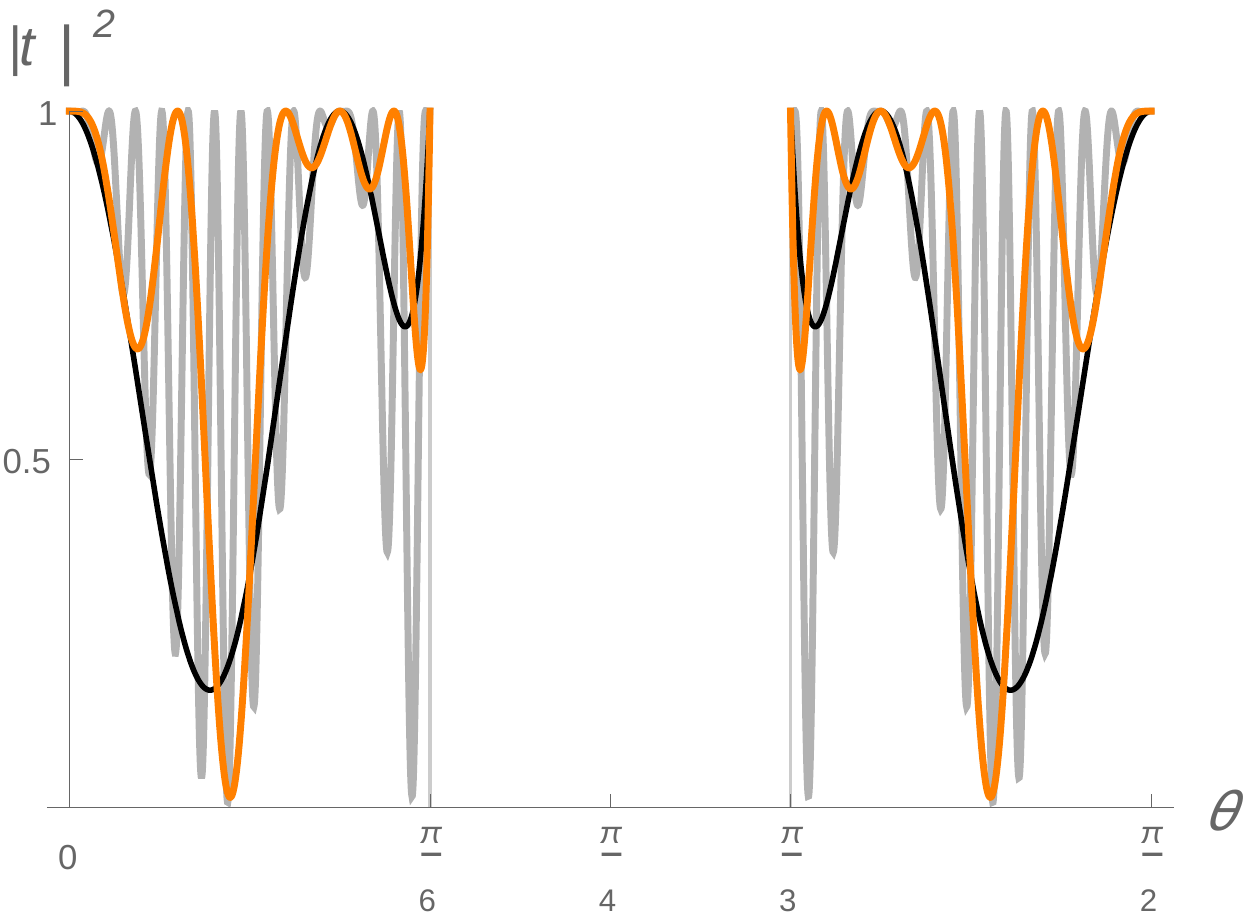} 
    \end{minipage}
    \hfill{
    \begin{minipage}{0.45\linewidth}
    \includegraphics[width=\linewidth]{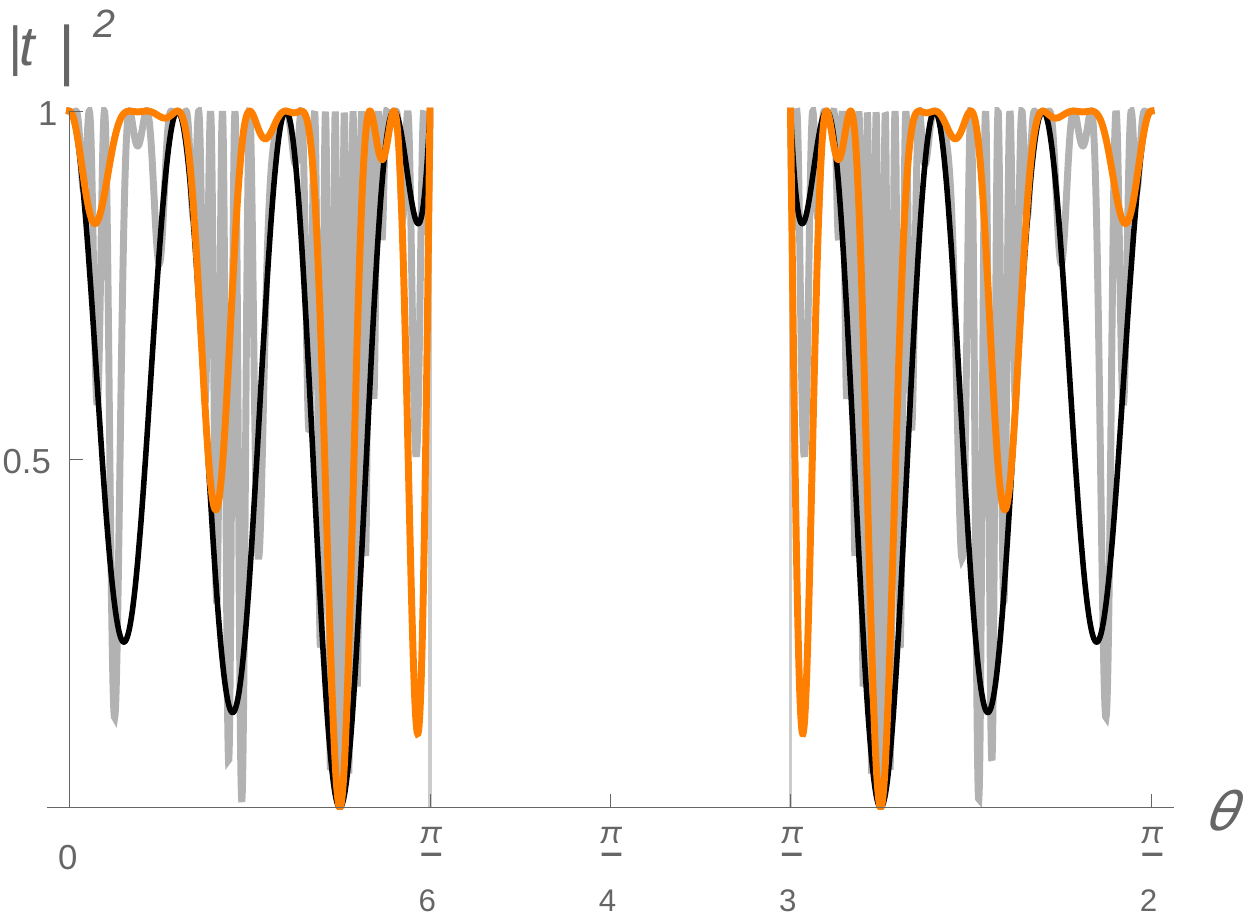}
    \end{minipage}
    }
    \caption{Transmission coefficient of the chain of $R_2^4$ blocks (left) with lengths $N=41$ (gray), $N=10$ (orange) and $N=1$ (black). Similar plot for $R_2^{10}$ blocks (right). Cases with $N=21$ (gray), $N=3$ (orange) and $N=1$ (black) are shown.}
    \label{fig:chain}
\end{figure}

This property breaks for $m\geq 4$, for which we can get non-trivial examples. In this case the transmission coefficients get extended areas of almost full transmission and peaks of low transmission, as demonstrated, for example, by figure~\ref{fig:chain} (left). As $m$ grows more regions with accumulations of minima appear, separated by gaps of high transmission, as in the $m=10$ example of fugure~\ref{fig:chain} (right). The loci of low transmission roughly correspond to minima of the $N=1$ case (torus knots and link).

Now we will consider two-block connection 
\begin{equation}
\label{twistknots}
t(\theta;l,n,m)=\langle\, \uparrow |R_2^l R_1^n R_2^m\up \,.
\end{equation}
and a particularly simple family of knots generated by such a connection, called twist knots.

\paragraph{Twist knots.} Twist knots are examples of knots with bridge number equal two~\cite{cromwell2004knots}. In terms of our construction this means that we can create them by acting once with the non-diagonal operator $R_2$ $(m=1)$ followed by $n$ applications of the operator $R_1^{-1}$ and finally $l=2$ applications of $R_2$. Therefore, the knots will differ only by the number $n$, and the number of crossings will be $n+2$ in each case, with the transmission coefficient 

\be 
|t(\theta;2,n,1)|^2 \ = \ 
|\langle\, \uparrow |(FR_1F^{-1})^2 (R_1^{-1})^n (FR_1 F^{-1})^1\up|^2
\ee

\begin{figure}[htb]
    \begin{minipage}{0.45\linewidth}
\includegraphics[width=\linewidth]{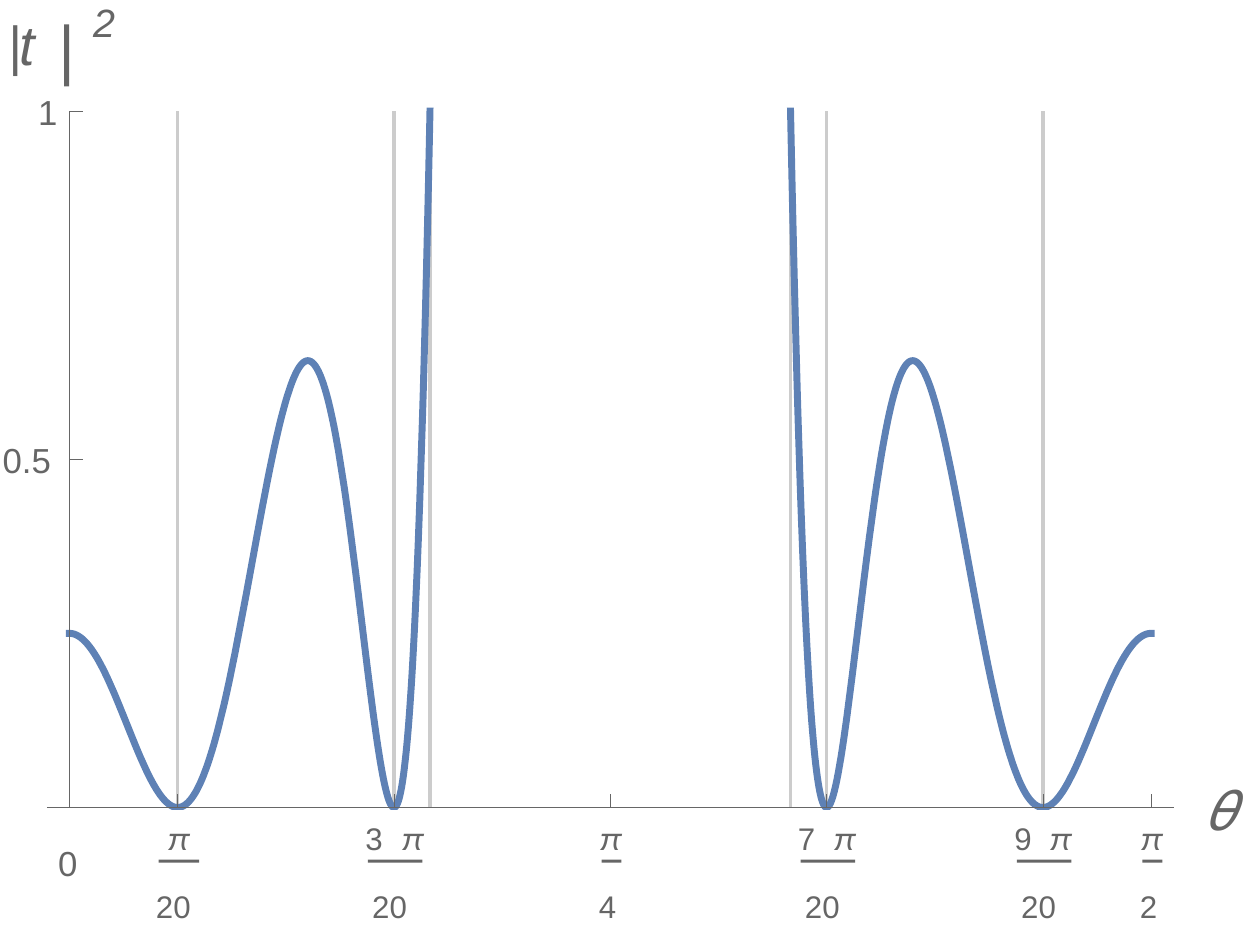}
    \end{minipage}
    {\hfill
    \begin{minipage}{0.45\linewidth}
\includegraphics[width=\linewidth]{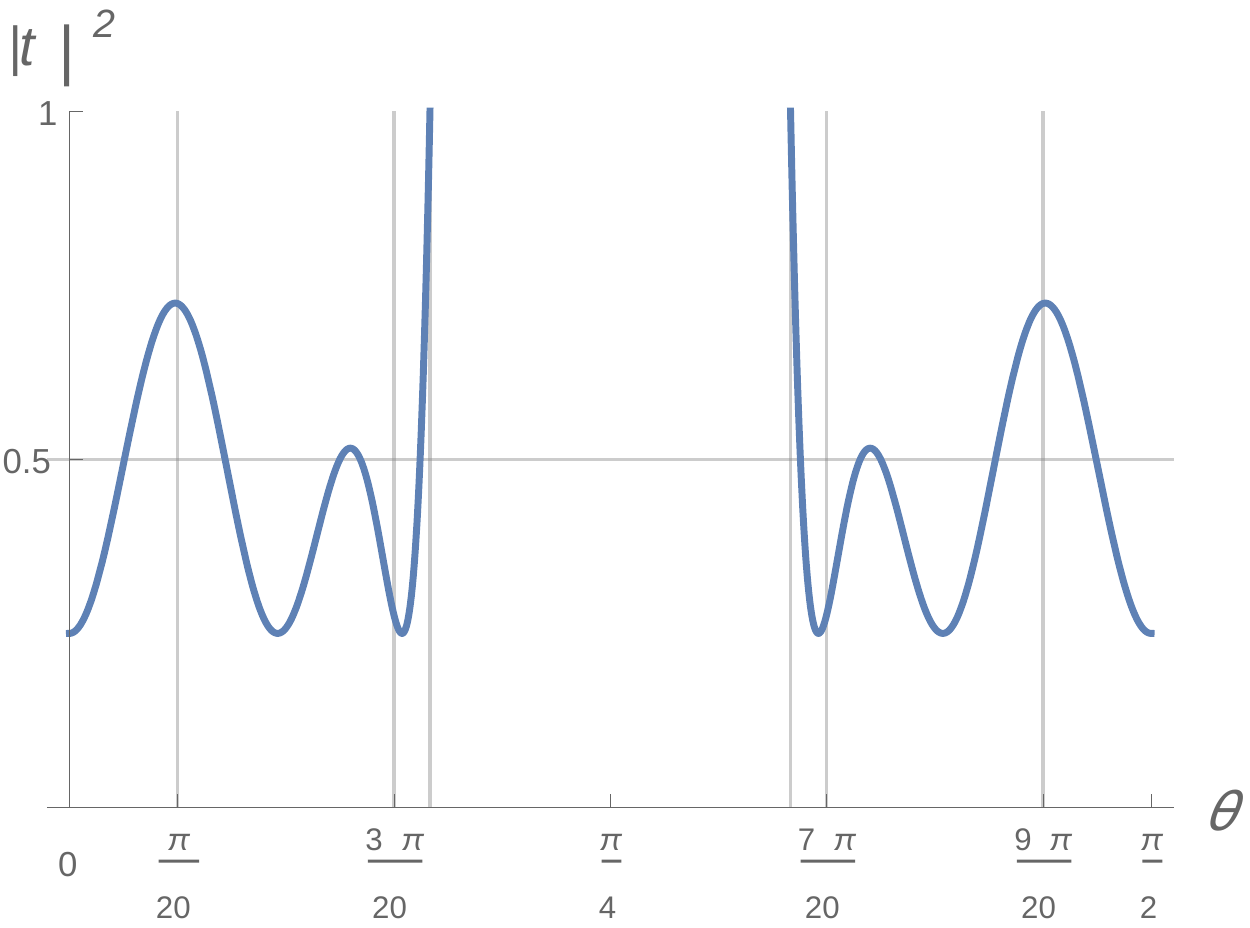}
    \end{minipage}
    }
    \caption{Transmission coefficient~(\ref{eqeight}) of the figure-eight knot $4_1$ (left). Transmission coefficient of $5_2$ knot (right).}
    \label{fig:t3}
\end{figure}

If $n=2$ one obtains the figure-eight knot, which is the simplest non-torus knot. Its transmission coefficient reads
\be 
|t(\theta;2,1,1)|^2 \ =\ |\langle\, \uparrow |(FR_1^2F^{-1})( R_1^{-1})( FR_1 F^{-1})\up|^2 \  = \ \frac{\sec^2{2\theta}}{4}(1-2\cos{4\theta}+2\cos{8\theta})^2\,.
\label{eqeight}
\ee
In comparison with the Hopf link and the trefoil knot the transmission coefficient of the figure eight has more zeroes and maxima in the unitary range, as can be seen in figure~\ref{fig:t3} (left). Function~(\ref{eqeight}) exhibits zeroes at $\theta=\{\frac{\pi}{20},\frac{3\pi}{20},\frac{7\pi}{20},\frac{9\pi}{20}\}\mod \pi/2$ and maxima at  $\theta=\pi/2$ ($|t|^2=1/4$), $\theta\sim 1.22$ $(|t|^2\sim 0.64)$, $\theta\sim 1.92$ $(|t|^2\sim 0.64)$, everything $\mod\pi/2$. The transmissions attains the maximum value only at the bound of unitarity, $\theta=\pi/6,\pi/3\mod\pi/2$.

As before, we can establish the relation of the transmission amplitude with the Jones polynomial,
\be
\label{eq:tjfig8}
|\langle\, \uparrow |R_2^2 R_1^{-1} R_2\up|\ = \ \frac{Jones(4_1)}{(-A^3)^2d}\,.
\ee

The next example in the family is the case $n=2$, which is the knot $5_2$ in the Rolfsen table. The transmission coefficient of this knot is given by 
\be 
|t(\theta;2,2,1)|^2 \ = \ |\langle\,\uparrow|R_2^2R_1^{-2}R_2 \up|^2\ =\ \frac{1}{4}\sec^2 2\theta[\cos^2 2\theta+(3\sin2\theta-2\sin6\theta+2\sin10\theta)^2]
\ee
whose plot is shown on figure~\ref{fig:t3}~(right). In terms of the Jones polynomial, the amplitude is
\be
\langle\,\uparrow|R_2^2R_1^{-2}R_2 \up \ = \ \frac{Jones(5_2)}{(-A^{-3})^5d}\,.
\ee
For this knot, the transmission is bound $1/4 \leq|t|^2\leq 3/4$, except for a region close to the unitarity bound. 

\begin{figure}[htb]
    \begin{minipage}{0.45\linewidth}
    \includegraphics[width=\linewidth]{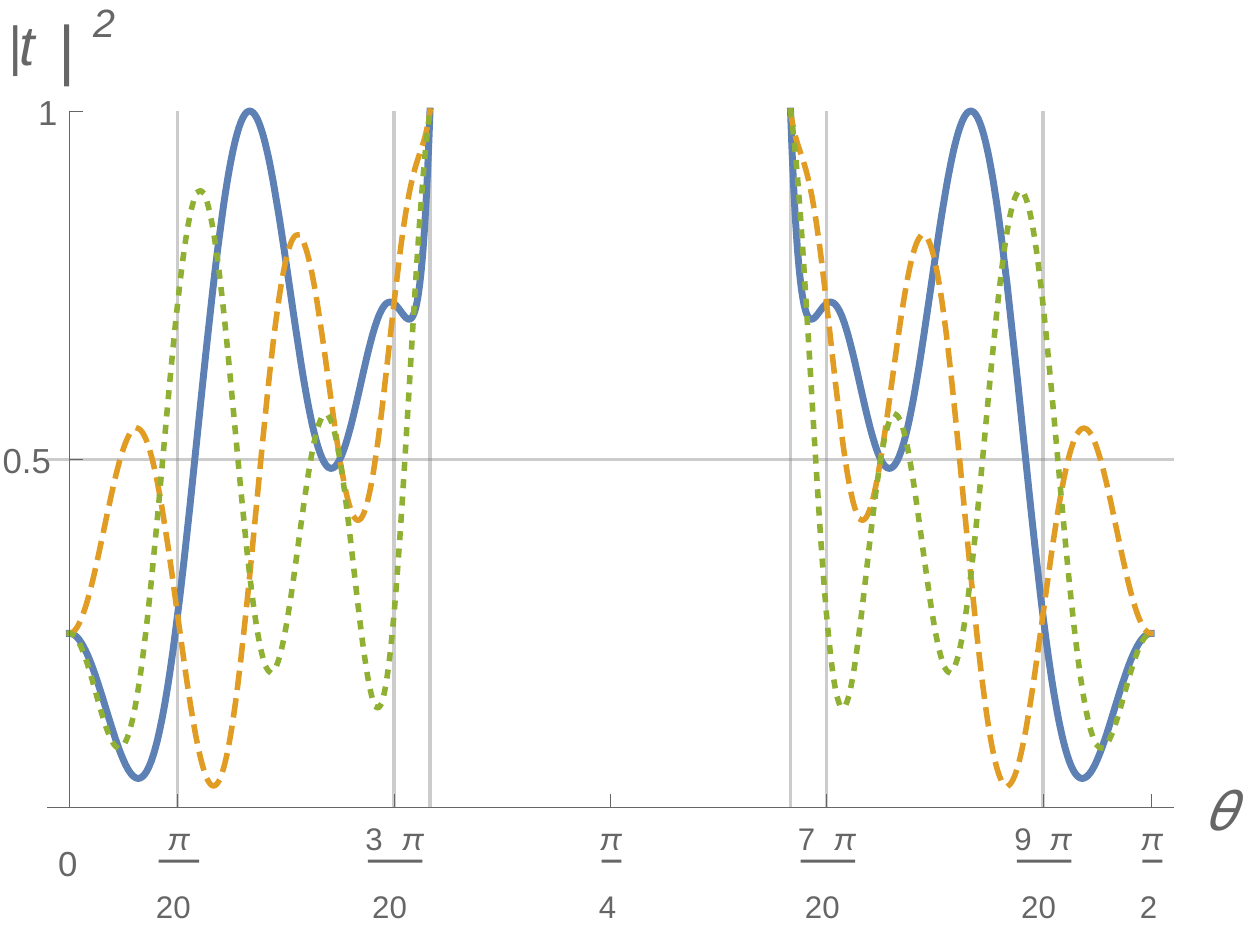}
    \end{minipage}
    \hfill{
    \begin{minipage}{0.45\linewidth}
    \includegraphics[width=\linewidth]{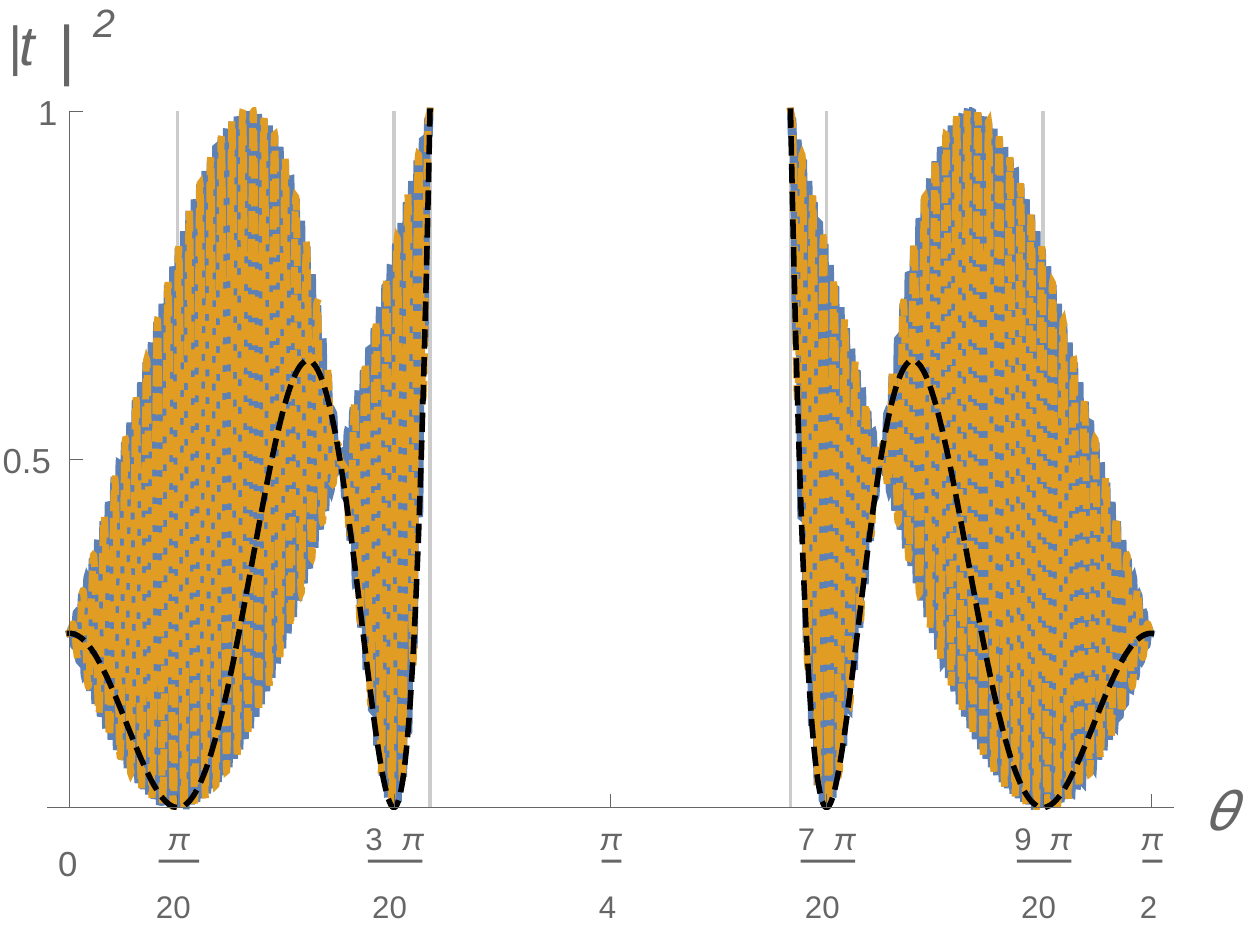}
    \end{minipage}
    }
    \caption{Transmission coefficients of $6_1$ (blue, solid), $7_2$ (orange, dashed) and $8_1$ (green, dotted) knots (left panel). Superposition of the plots for $n=100$ and $n=200$, illustrating the envelope curves for the twist family (right panel). Plots of the coefficient for the figure-eight knot is shown by the black dashed curve.}
    \label{fig:twistfamily}
\end{figure}

We also summarize the coefficients of a few other representatives (see left plot of figure~\ref{fig:twistfamily}),
\begin{itemize}
    \item $6_1$ knot, $n=3$
    \be
    \langle\,\uparrow|R_2^2R_1^{-3}R_2 \up \ = \ - \frac{Jones(6_1)}{(-A^{-3})^2d}\,,
    \ee
    \be 
    |t(\theta;2,3,1)|^2 \ = \ \frac{1}{4}\sec^2 (2\theta)[\sin^2 2\theta+(-2+3\cos4\theta-2\sin8\theta+2\sin12\theta)^2]\,;
    \ee
    \item $7_2$ knot, $n=4$,
    \be
    \langle\,\uparrow|R_2^2R_1^{-4}R_2 \up \ = \ \frac{Jones(7_2)}{(-A^{-3})^7d}\,,
    \ee
    \be 
    |t(\theta;2,4,1)|^2 \ =\ \frac{1}{4} \sec ^2(2 \theta) \left((-4 \sin (2 \theta)+3 \sin (6 \theta)-2 \sin (10 \theta)+2 \sin (14 \theta))^2+\cos^2(6 \theta)\right)\,;
   \ee
    
    \item $8_1$ knot, $n=5$,
    \be
    \langle\,\uparrow|R_2^2R_1^{-5}R_2 \up \ =\ -\frac{Jones(7_2)}{(-A^{-3})^4d}\,,
    \ee
    \be 
    |t(\theta;2,5,1)|^2 \ =\ \frac{1}{4} \sec ^2(2 \theta) \left((-4 \cos (4 \theta)+3 \cos (8 \theta)-2 \cos (12 \theta)+2 \cos (16 \theta)+2)^2-\sin^2(8 \theta)\right)\,;
   \ee
    
    \item general $n$, 
    \begin{multline}
    |\langle\,\uparrow|R_2^2R_1^{-n}R_2 \up|^2 \ = \ \frac{1}{8\cos^42\theta}\bigg(2+\cos8\theta-\cos12\theta + \\ + (-1)^n(\cos(4(n-1)\theta)+\cos(4(n+1)\theta)-\cos(4(n+2)\theta)-\cos(4(n+4)\theta))\bigg)\,.
    \end{multline}
    For $n=1\mod 5$ the transmission coefficients have exact zeroes, at the position of the zeroes of the figure-eight. Some representatives of the family also exhibit maxima of full transmission. See~\cite{Mironov:2014zza} for general results for the invariants of the twist knots.

\end{itemize}

The structure of the transmission coefficients of this family is similar to the structure of that of torus knots. Higher representatives of the family produce higher order polynomials, which have higher numbers of minima and maxima. These minima and maxima are bound by two envelope curves for this family. In comparison with the torus family, the upper envelope has a shape different from constant, as shown in figure~\ref{fig:twistfamily}~(right). An interesting property of this family is the point $\theta= \pi/8$ where two envelopes intersect at half of the full transmission. This corresponds to a universal value $-1$ of the Jones polynomial of the twist family.

\paragraph{Generalized twist family.} Finally, we consider a few representative of the generalized twist family, which is also called \emph{double braid knots}~\cite{Mironov:2014eza,Morozov:2016eqp}. This family can be obtained by keeping arbitrary $l=1$, $m>2$ and $n$ in equation~(\ref{twistknots}). We choose $m=3$ in the explicit examples:
\be 
|t(\theta;1,n,3)|^2\ =\ |\langle\, \uparrow |(FR_1F^{-1})^1 (R_1^{-1})^n (FR_1 F^{-1})^3\up|^2
\ee 

\begin{itemize}
\item $7_3$ knot, $n=3$:
\be
\langle\,\uparrow|R_2R_1^{-3}R_2^3 \up \ = \ \frac{Jones(7_3)}{(-A^3)^7d}
\ee
\be |t(\theta;1,3,3)|^2 \ =\ \frac{1}{4} \sec ^2(2 \theta) \left(\cos ^2(2 \theta)+(-5 \sin (2 \theta)+4 \sin (6 \theta)-2 \sin (10 \theta)+2 \sin (14\theta))^2\right)
   \ee
\item $9_3$ knot, $n=5$:
\be
\langle\,\uparrow|R_2R_1^{-5}R_2^3 \up \ = \ \frac{Jones(9_3)}{(-A^3)^9d}
\ee
\begin{multline}
|t(\theta;1,5,3)|^2 \\ = \ \frac{1}{4} \sec ^2(2 \theta) \left((6 \sin (2 \theta)-5 \sin (6 \theta)+4 \sin (10 \theta)-2 \sin (14 \theta)+2 \sin (18 \theta))^2+\cos ^2(6 \theta)\right)    
\end{multline} 

\item general $n$:
\be
\label{twistamps}
\langle\,\uparrow|R_2R_1^{-n}R_2^3 \up \ = \ \frac{A^{-4n-8}(1+A^8+A^{16})+(-1)^n(-1+A^4+A^{12})}{A^{3n}(1+A^4)^2}\,.
\ee

\item general $n$ and general $m$.
\begin{multline}
\langle\,\uparrow|R_2R_1^{-n}R_2^{m} \up \\ = \ \frac{(A^{1+m-4n}-(-1)^{m}A^{1-3m-4n}+(-1)^nA^{5+m})(1+A^4+A^{8})+(-1)^{n+m}A^{9-3m}}{A^{3n}(1+A^4)^3}\,.
\end{multline}
This can be compared with general formulae for the Jones polynomials of the twist knots~\cite{Morozov:2016eqp}.
\end{itemize}

\begin{figure}[htb]
    \begin{minipage}{0.45\linewidth}
    \includegraphics[width=\linewidth]{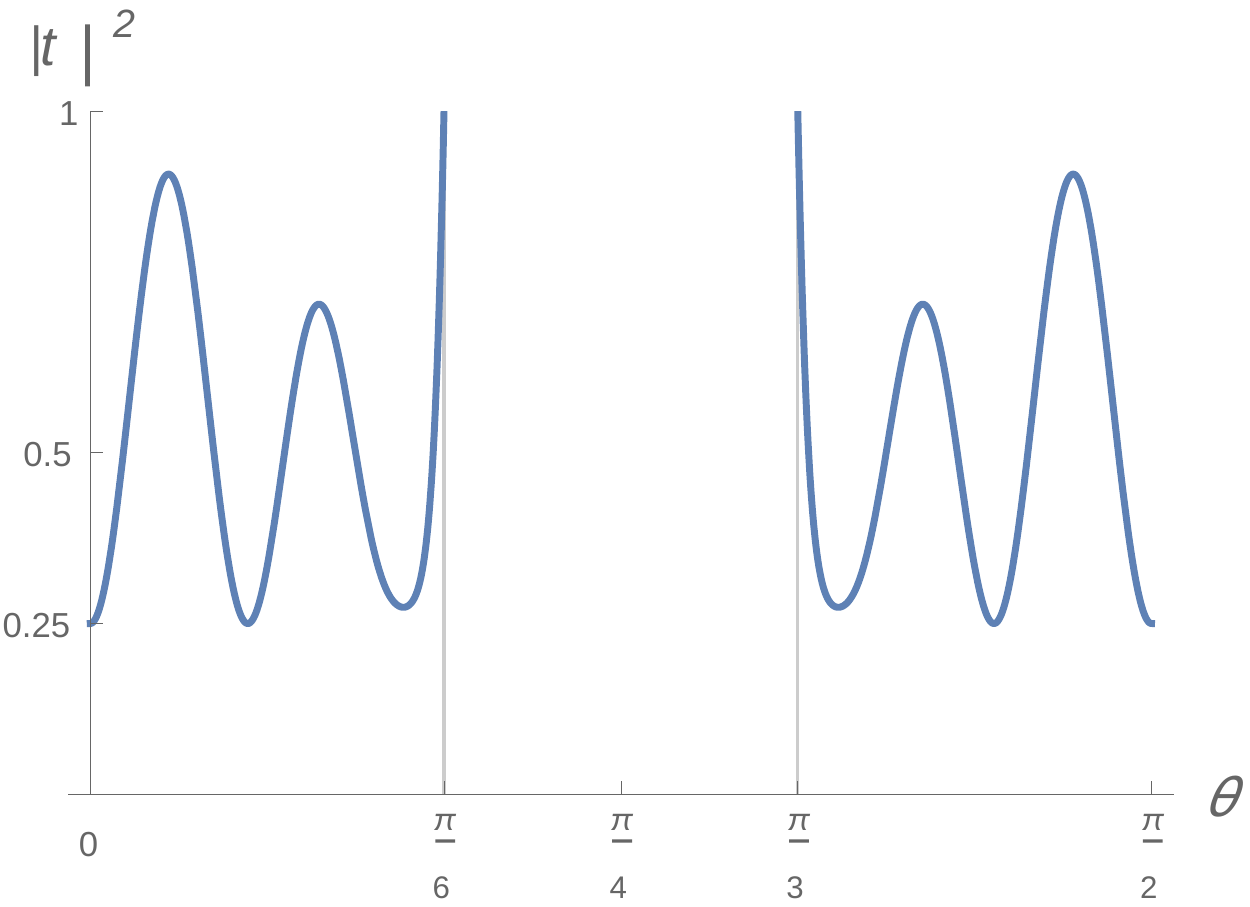}
    \end{minipage}
    \hfill{
    \begin{minipage}{0.45\linewidth}
    \includegraphics[width=\linewidth]{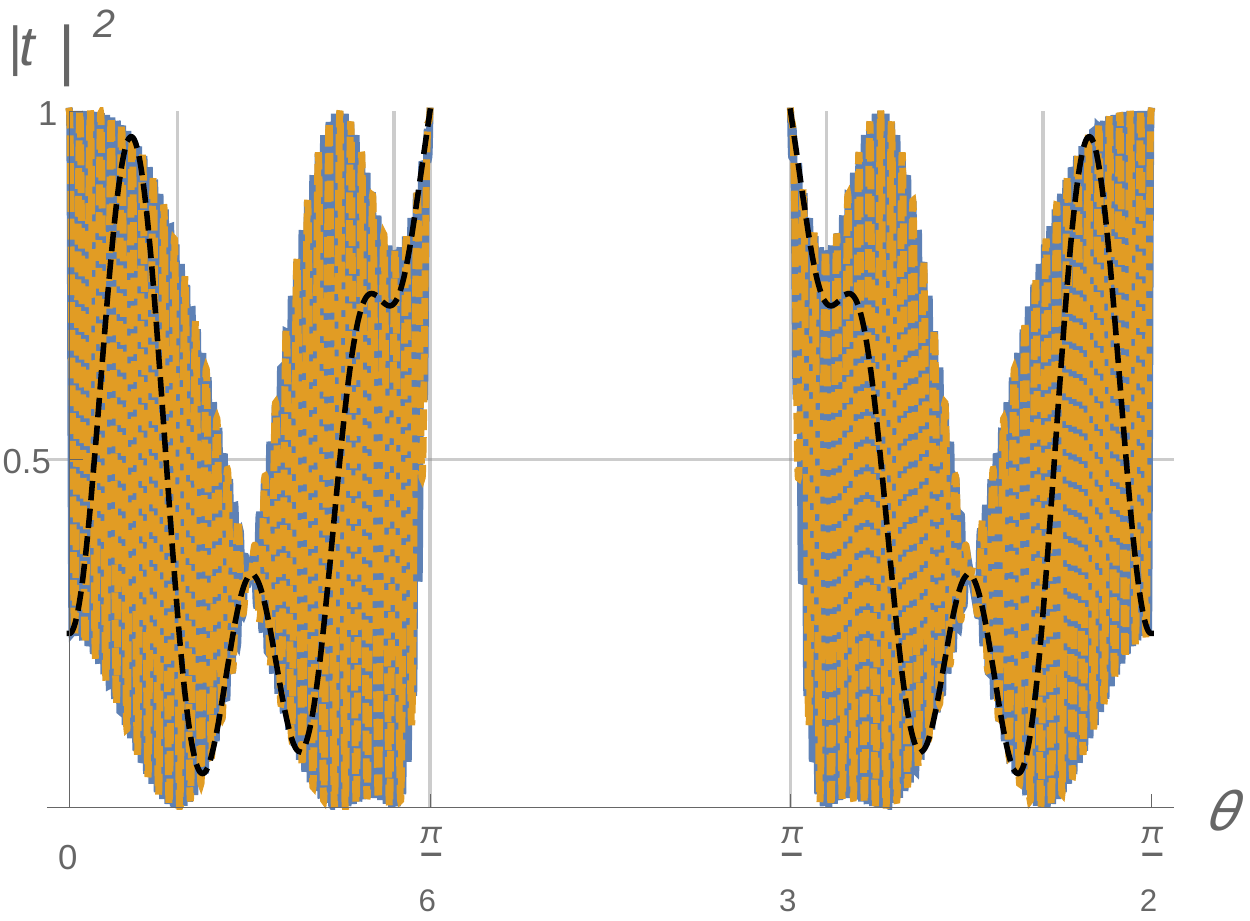}
    \end{minipage}
    }
    \caption{Transmission coefficient of $7_3$ knot (left). Transmission coefficient of $9_3$ knot (black, dashed) superposed on large $n$ limit of the transmission coefficient (right).}
    \label{fig:gentwist}
\end{figure}

The shape of the transmission coefficient for two representatives of the $m=3$ double braid family is illustrated in figure~\ref{fig:gentwist}. For this family, the universal point is shifted to $\theta={\pi}/12$ with the value of the transmission coefficient $|t|^2=1/3$.
   
One can also find the universal points for higher representatives. For $m=4$, $\theta_u={\pi}/8$ and $|t|^2=1/2$, as for the $m=2$ family. For $m=5$, $\theta_u=\pi/20$ with $|t|^2=(3+\sqrt{5})/(10+4\sqrt{5})\sim 1/4$.

%%%%%%%%%%%%%%%%%%%%%%%%%%%%%%%%%%%%%%%%%%%%%%%%%%%%%%%%%%%%%%%  

\section{Pretzel knots}
\label{sec:pretzel}

In this section we consider a generalization of the model to examples of ``multichannel" transmission. The motivation for the particular generalization to be considered is a class of ``pretzel" knots and links considered in~\cite{Galakhov:2014sha,Mironov:2014eza,Galakhov:2015fna}.  The pretzel diagrams for the knots of this family provide a natural generalization of the diagrams considered above to multiple channels. Pretzel knots themselves can be considered as a generalization of two strand torus knots to genus $g$ Riemann surfaces (they can be drawn on the Riemann surface without self-intersections). Their invariants are also given by a closed-form expression.\footnote{For general information on pretzel knots see~\cite{cromwell2004knots}.}

\begin{figure}[htb]
    \begin{minipage}{0.45\linewidth}
    \includegraphics[width=\linewidth]{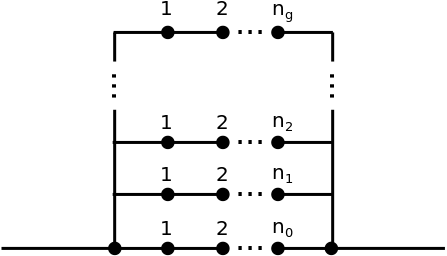}
    \end{minipage}
    \hfill{
    \begin{minipage}{0.45\linewidth}
    \includegraphics[width=\linewidth]{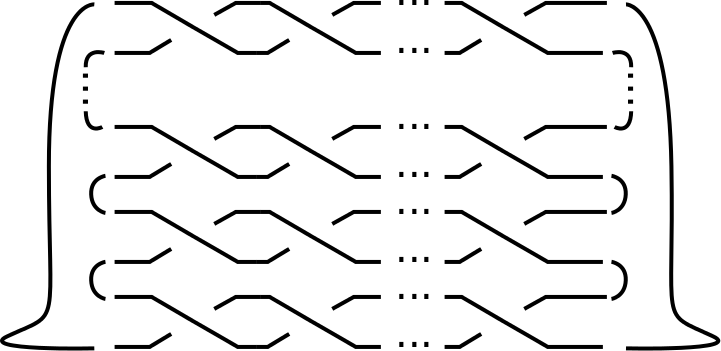}
    \end{minipage}
    }
    \caption{Multichannel transmission graph (left) and its possible resolution in terms the pretzel knots of~\cite{Galakhov:2014sha,Mironov:2014eza} on genus $g$ Riemann surfaces (right).}
    \label{fig:pretzel}
\end{figure}

In figure~\ref{fig:pretzel} we show a possible graph construction realizing a multichannel transmission and a resolution of the graph in terms of a link diagram. Note, as before, the resolution does not distinguish the position of the leads along the vertical segments of the graph.

As can be seen from the figure, pretzel knots are defined by the set of integer numbers $\{n_0,n_1,\ldots,n_g\}$ -- the number of braidings $n_i$ in each of the $g+1$ channels. As one can learn from~\cite{Galakhov:2014sha} the (unnormalized) Jones polynomials of the pretzel knots, can be calculated as
\be
Jones(n_0,n_1,\ldots,n_g) \ = \ \frac{(-1)^{g+1}}{[2]^{g+2}}\left(\prod\limits_{i=0}^g(1+[3](-q)^{n_i})+[3]\prod\limits_{i=0}^g(1-(-q)^{n_i})\right)\,,
\ee
where $[n]$ denotes the ``quantum numbers",
\be
[n] \ = \ \frac{q^{n/2}-q^{-n/2}}{q^{1/2}-q^{-1/2}}\,.
\ee
Identifying $A\equiv q^4$ and normalizing appropriately one can compute the transmission amplitudes. 

Note that the canonical normalization of the Jones polynomials corresponds to unity for the unknot up to a phase (framing factor). For a collection of disconnected loops this normalization yields the factor of $[2]^{L-1}$, where $L$ is the number of loops and $[2]=-d$ corresponds to the quantum dimension of the fundamental representation of $SU(2)$. For this reason the factor of $d$ appeared in the relation between the Jones polynomials and amplitudes like equations~(\ref{torusamps}) and~(\ref{twistamps}). It cancelled the extra factor of $d$, since the most basic diagram with no braiding contains two loops. As the result all amplitudes were unitary in the region~(\ref{eq: domain}).

Higher genus diagrams would bring extra factors of $d$ in the numerator, for example,
\be
Jones(0,0,\ldots,0) \ = \ d^g\,.
\ee
This creates a problem for unitarity for $g>1$.

In the multichannel problem the representations of the genus $g$ braid group $B_{g+1}$ are constructed as tensor products of $B_3$ and we expect these representations to be unitary. Therefore, the problems with the unitarity arise at the level of the normalization of the initial and final states. It is natural to associate the trivial braid group element to a process with no scattering. From this point of view, the transmission coefficient is given by
\be
\label{tg}
|t_g(n_0,\ldots,n_g)|^2 \ = \ \frac{|Jones(n_0,\ldots,n_g)|^2}{d^{2g}}\,.
\ee

However, we are not interested in the scattering of the states in a bigger Hilbert space, but rather of the states like $\up$ discussed before. In such a case we should keep the original normalization and calculate the transmission coefficient as
\be
\label{t2}
|t(n_0,\ldots,n_g)|^2 \ = \ \frac{|Jones(n_0,\ldots,n_g)|^2}{d^{2}}\,.
\ee
In this case the procedure of encoding the original states in the bigger Hilbert space is not always unitary even in the region defined by~(\ref{eq: domain}). This can be seen for $g>1$ and $n_i>1$ for all $i$, for example. One the other hand the transmission coefficient remains a knot invariant as equation~(\ref{t2}) implies (as opposed to (\ref{tg}) which depends on the way the knot is drawn).
 
\begin{figure}[htb]
    \begin{minipage}{0.45\linewidth}
    \includegraphics[width=\linewidth]{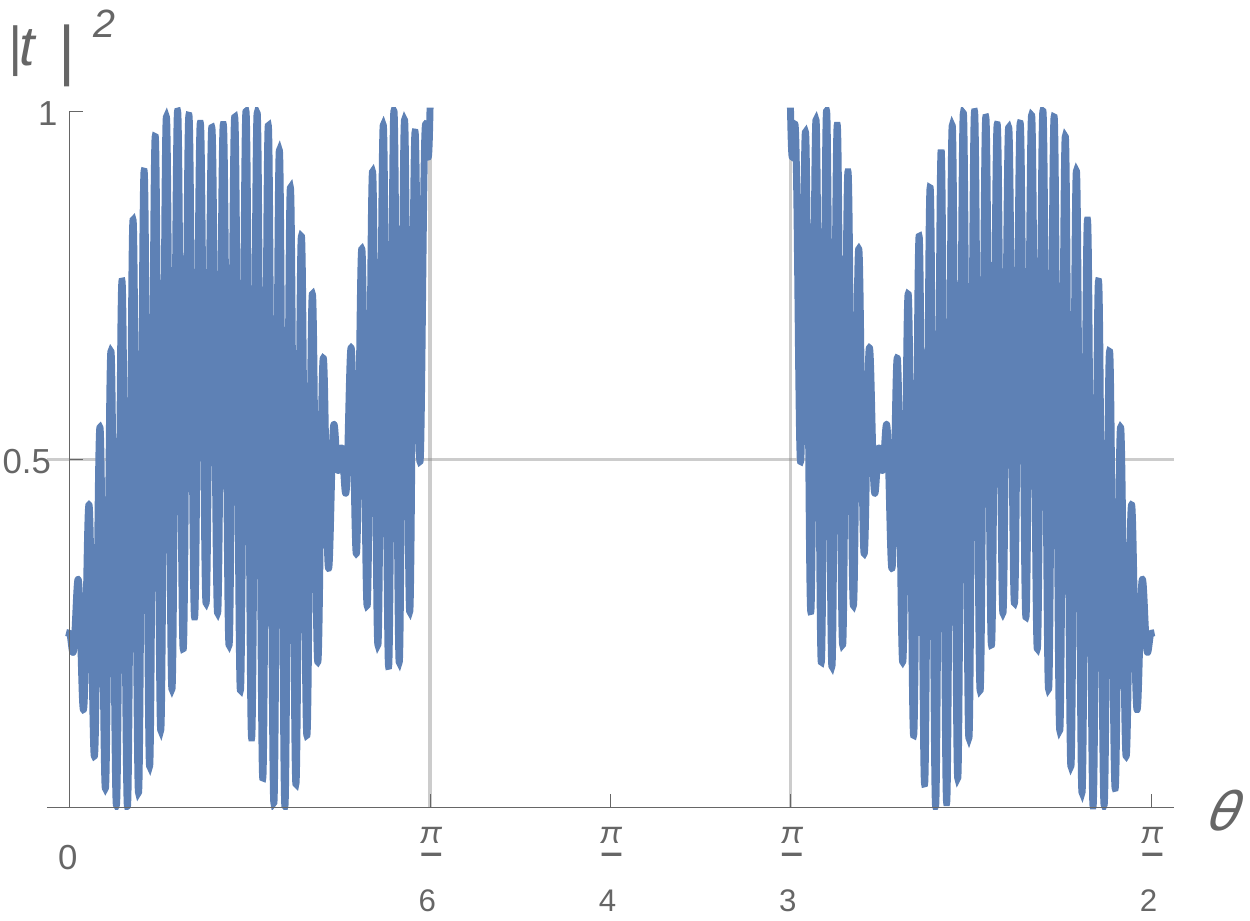}
    \end{minipage}
    \hfill{
    \begin{minipage}{0.45\linewidth}
    \includegraphics[width=\linewidth]{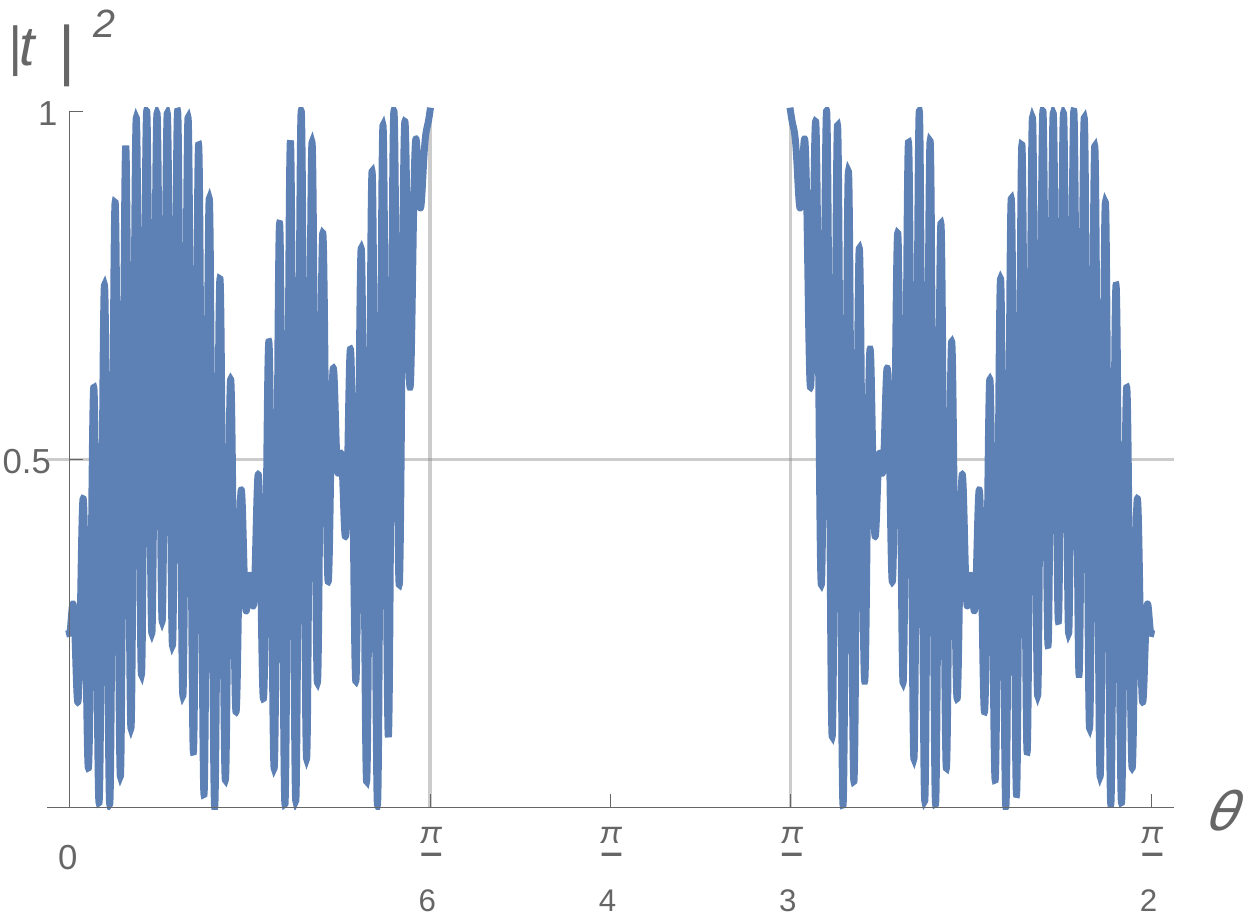}
    \end{minipage}
    }
    \caption{The transmission coefficients of $(-100,1,-5)$ and $(-100,1,5)$ pretzel knots.}
    \label{fig:multichannel}
\end{figure}

A few examples of the transmission coefficients is shown in figure~\ref{fig:multichannel}. As the figure illustrates, variation of the numbers $n_i$ allows further modulation of the oscillations, producing envelopes with additional features, as compared to the simple families considered before. In particular one can construct families with more universal points in the interior of the unitary interval.

%%%%%%%%%%%%%%%%%%%%%%%%%%%%%%%%%%%%%%%%%%%%%%%%%%%%%%%%%%%%%%%
   
\section{Physical realization}
\label{sec:physical}

In this section we would like to place the topological model of transmission in an appropriate physical context. The most pertinent systems, where such transmission can be discussed are the quantum Hall systems, or quantum spin Hall systems~\cite{DasSarma:2005zz,Nayak:2008zza}. Quantum Hall effect arises in two-dimensional conductors subject to strong magnetic fields. Magnetic field create highly degenerate energy (Landau) levels separated by gaps proportional to the value of the field. As a result, the bulk of the system behaves as an insulator if the Fermi energy is tuned to be between the Landau levels.

Boundary effects change the structure of the energy levels. Their degeneracy is lifted and the gaps are closed, providing a massless (edge) mode for each level below the Fermi energy. The edge modes contribute dissipationless edge currents in the direction prescribed by the orientation of the magnetic field.

In terms of the diagrams in figure~\ref{fig:network1}, from which we introduced the model, the edges of the graphs can be viewed as loci of the currents. Then the face of the diamond in the figure is the bulk of a quantum Hall material and the external edges are leads. The nodes of the diagram can be interpreted as interactions externally induced on the edge currents. The flow of the edge current is then described by a knot diagram. 

Due to their dissipationless nature the edge currents do not depend on the shape and other details of the edge. This topological property is a defining characteristic of the knot diagrams. Moreover, quantum Hall effect is effectively described by 2+1 dimensional Chern-Simons theory, a topological gauge theory. In the presence of an edge, the Chern-Simons theory is supplemented by edge terms, which are typically in the form of a 1+1 dimensional chiral CFT, describing the edge modes. The specific choice of the CFT (equivalently, Chern Simons theory) defines the type of interactions available for the edge modes and their currents. They are encoded in the chiral algebra of the CFT.

The chiral algebra defines the fusion and the braiding properties of the CFT operators, which are reflected upon the edge states in the form of the action of $R$ and $F$ matrices discussed above. Therefore we can relate the main parameter $\theta$ of our model with the parameters of Chern-Simons and CFT theories. For a non-abelian, $SU(2)$ Chern-Simons, whose obervables are precisely described in terms of the Jones polynomials~\cite{Witten:1988hf}, this connection would be
\be
\theta \ = \  \frac{\pi}{k+2}\,,
\ee
where $k$ is the level (the coupling constant) of the Chern-Simons theory. In particular, $k\to\infty$ and $\theta\to 0$ is the classical limit of the Chern-Simons theory. The unitarity constraints that appeared before, simply mean that $k\geq 4$, or $k\leq 1$. The second range corresponds to the set of electric-magnetic (particle-vortex) dual theories related by $k\to 1/k$. 

One possible experimental realization of the torus family~(\ref{HopfBraid2}) can be described via the setup of~\cite{barkeshli2014synthetic}, where a concept of tunneling interaction between two edge currents was proposed in a bilayer Hall bar configuration. A more simple example is a quantum Hall bar with a constriction (see~\cite{Nayak:2008zza} and reference therein), where tunneling is induced between the opposite edges of a quantum Hall bar. The latter example is analogous to the configuration of the twist family, like in equation~(\ref{twistnetwork}). In either case, we can model this basic tunneling effects with the braiding operations in the topological model.

\section{Conclusions}
\label{sec:conclusions}

In this work we considered realizations of elementary quantum scattering in topological quantum field theories. We used unitary representations of the braid group to connect knots and links with the evolution of quantum states in such theories. We showed how different patterns of transmission can be obtained choosing different knots.

We argued that such scattering processes can be realized in quantum Hall systems. In particular, the knot diagrams can describe the configurations of the edge charge or spin currents. The simplest experiments, in which a quantum Hall states is realized on a bar of simple topology, correspond to the families of $(n,2)$ torus and twist knots in the topological model. More complicated multichannel scattering processes can be described in terms of the pretzel knots.

An interesting feature of the scattering is the existence of universal ``self-averaging" points in the phase diagram. These correspond to the universal values of the Jones polynomials at roots of unity. For the twist family there is a single universal point in the interior of the unitary domain (the boundaries of the domain always have a universal transmission withing the classes of knots or links). More self-averaging points can be created in the multichannel scattering. Such points should play a prominent role in the models with random scattering.

In our study we used a simple scattering model on graphs as a prototype. We observed that the scattering problem in the graph and in the topological models reduce to solving a set of linear recurrence relations (skein relations). It would be interesting to expand the analogy further and to construct a graph model based on the skein relations of knots.

\paragraph{Acknowledgements.} The author is grateful to Dion\'isio Bazeia for explaining the results of his earlier work and for the fruitful discussions, from which this project was conceived. He would also like to thank Andrey Morozov and Alexey Sleptsov for useful comments and Victoria Maria Leite for her collaboration on parts of this project. This work was supported by the Russian Science Foundation grant No. 18-71-10073.

\bibliographystyle{hieeetr}
\bibliography{refs}
\end{document}